\RequirePackage{fix-cm}
\documentclass[twocolumn]{svjour3}          % twocolumn
\smartqed  % flush right qed marks, e.g. at end of proof
\usepackage{graphicx}
\usepackage[colorinlistoftodos]{todonotes}
\usepackage{multirow}
\usepackage{hyperref}
\usepackage{array}
\usepackage{tabu}
\usepackage{tabularx}
\newcolumntype{M}[1]{>{\centering\arraybackslash}m{#1}}
\newcolumntype{R}[1]{>{\raggedleft\arraybackslash}m{#1}}
\newcolumntype{L}[1]{>{\arraybackslash}m{#1}}

\begin{document}
\sloppy

\title{What is all this new MeSH about?
}
\subtitle{Exploring the semantic provenance of new descriptors in the MeSH thesaurus
% Where do new MeSH descriptors come from?
}


\author{Anastasios Nentidis$^{1,2}$ \and Anastasia Krithara$^1$ \and Grigorios Tsoumakas$^2$ \and Georgios Paliouras$^1$}

% \author{First Author         \and
%         Second Author %etc.
% }

\authorrunning{Nentidis \textit{et al.}} % if too long for running head

% \institute{F. Author \at
%               first address \\
%               Tel.: +123-45-678910\\
%               Fax: +123-45-678910\\
%               \email{fauthor@example.com}           %  \\
% %             \emph{Present address:} of F. Author  %  if needed
%               second address
% }

\institute{
$^1$National Center for Scientific Research ``Demokritos'', Athens, Greece \\
\email{\{tasosnent, akrithara, paliourg\}@iit.demokritos.gr}\\
  $^2$Aristotle University of Thessaloniki, Thessaloniki, Greece\\
\email{\{nentidis, greg\}@csd.auth.gr}\\
}
\date{Received: date / Accepted: date}

\maketitle

\begin{abstract}
% Structured abstract: Please provide a structured abstract of 150 to 250 words which should be divided into the following sections: Purpose (stating the main purposes and research question) Methods Results Conclusion
% purpose
The Medical Subject Headings (MeSH) thesaurus is a controlled vocabulary widely used in biomedical knowledge systems, particularly for semantic indexing of scientific literature.
As the MeSH hierarchy evolves through annual version updates, some new descriptors are introduced that were not previously available.
This paper explores the conceptual provenance of these new descriptors. 
In particular, we investigate whether such new descriptors have been previously covered by older descriptors and what is their current relation to them.
% methods
To this end, we propose a framework to categorize new descriptors based on their current relation to older descriptors. 
Based on the proposed classification scheme, we quantify, analyse and present the different types of new descriptors introduced in MeSH during the last fifteen years. 
% results
The results show that only about 25\% of new MeSH descriptors correspond to new emerging concepts, whereas the rest were previously covered by one or more existing descriptors, either implicitly or explicitly. Most of them were covered by a single existing descriptor and they usually end up as descendants of it in the current hierarchy, gradually leading towards a more fine-grained MeSH vocabulary. 
% conclusion
These insights about the dynamics of the thesaurus are useful for the retrospective study of scientific articles annotated with MeSH, but could also be used to inform the policy of updating the thesaurus in the future.

\end{abstract}

\section{Introduction}
\label{intro}
% \todo[inline]{Introduce MeSH.}
The \textit{Medical Subject Headings} (MeSH) thesaurus\footnote{\url{https://meshb.nlm.nih.gov/}} is a collection of hierarchically organized entities for annotating biomedical knowledge, primarily literature in PubMed/MEDLINE\footnote{\url{https://www.nlm.nih.gov/bsd/pmresources.html}}, with topic labels. The basic conceptual entity is the MeSH \textit{concept} which is a collection of synonymous terms for a particular domain meaning. Each concept has a {\em preferred term}, which is also used as the name of the concept. 
MeSH concepts are not directly used for annotating the literature. Instead, closely related concepts are grouped into MeSH {\em descriptors} that constitute the main MeSH elements for annotating biomedical literature with topic labels. Although a descriptor can consist of several concepts, each MeSH concept belongs to exactly one MeSH descriptor.
All the concepts and terms of a descriptor are equivalent for the purposes of indexing and searching MEDLINE.
Beyond MeSH concepts and descriptors,
MeSH also provides some \textit{Supplementary Concept Records} (SCRs) that are directly used for annotating articles with labels for substances, rare diseases, and organisms.\footnote{\url{https://www.nlm.nih.gov/mesh/intro_record_types.html}}

As the MeSH hierarchy evolves through annual updates, new descriptors are introduced that were previously unavailable. 
This evolution of MeSH is essential, in order to follow the development of knowledge in the field. For example, new descriptors can be more fine-grained than old ones, providing a level of detail previously unavailable in the vocabulary.
On the other hand, new high-level descriptors can also be added, providing new groupings of topics, under the light of the current understanding of the domain.
In some cases, the topics covered by the new descriptors may have been present in MeSH previously, covered by older descriptors.
However, some new descriptors may cover topics that are totally new to the vocabulary, representing emerging concepts in the domain.

Despite their necessity for keeping MeSH up-to-date, the introduction of new descriptors raises practical challenges. Several applications, such as (semi-)automated semantic indexing of biomedical literature with MeSH labels, are based on supervised learning techniques that exploit accumulated data from previous use of the vocabulary. However, for new descriptors, no such annotated literature is available at the time of their introduction.
Therefore, it becomes important to devise a mapping of existing literature to the new descriptors.
Towards this direction, for each new descriptor, we are interested in whether the corresponding topic was already covered by old descriptors in MeSH or not. 
% Different descriptors can have different $year_1$ and $year_0$. 

In this context, the basic questions motivating this study on the provenance of new descriptors are the following:
\begin{itemize}
    \item To what extent do the new MeSH descriptors cover emerging domain concepts that are really new for the MeSH thesaurus? 
    \item 
    % Are there any older descriptors that were previously used to index articles belonging to the topics of new descriptors? How can they be identified?
    For those new descriptors that do not cover emerging domain concepts, can we identify older descriptors that were used to cover these concepts?
    \item 
    % How do the new descriptors currently relate to any old ones previously used for indexing the same topic? Do the latter still cover the same concepts?
    What is the current relation of the new descriptors with the old ones that they are related to?
    \item 
    % How consistent are the conceptual-provenance dynamics outlined by the above questions during the last years in MeSH?
    Is there any pattern over time concerning the introduction of new descriptors in MeSH and how the new descriptors relate to the old ones?
\end{itemize}

The main contribution of this work consists in developing a conceptual framework for exploring the provenance of new MeSH descriptors considering the hierarchical structure of the thesaurus. 
In particular, we describe an approach for identifying predecessor descriptors, that used to cover the topic of a new descriptor previously. Namely, a coding system is introduced for organizing the new descriptors based on two key dimensions: a) whether and how they have been covered in MeSH prior to their introduction, and b) their current position in the hierarchy in relation to their predecessors.  
In addition, a method is developed for the computational identification of predecessors and conceptual provenance codes for new MeSH descriptors.
Finally, based on the proposed framework we perform an analysis that sheds light on the conceptual provenance of descriptors introduced in MeSH during the last fifteen years.

The rest of this paper is structured as follows. 
In Section~\ref{intro_mesh_structure} some background knowledge is summarized, regarding the structure of basic elements of the MeSH thesaurus and their relationships.   
In section~\ref{sec:RelatedWork} we provide a brief overview of work related to the extension of biomedical thesauri with new concepts, focusing on the MeSH thesaurus. 
In section~\ref{sec:ConceptualFramework} we propose a framework for identifying the predecessors of new descriptors and introduce new types of conceptual provenance to characterize the current relation of new MeSH descriptors with their predecessors. 
In section~\ref{sec:DataAnalysis} we propose a method to automatically analyze the versions of the MeSH hierarchy, in order to identify the various types of provenance.
In section~\ref{sec:results} we present and discuss the results of this analysis, which lead to useful insights about the evolution of MeSH. Finally, in section~\ref{sec:Conclusion} we conclude and indicate potential uses of our results in future research.

\section{MeSH structure}
\label{intro_mesh_structure}

The MeSH hierarchy is elaborately structured to efficiently represent and organize terms, concepts, and topics from the complex domain of biomedicine.  
% is a comprehensive terminological resource organizing topics and terms about complex biomedical phenomena. As a result, it 
This elaborate structure of MeSH is manually maintained by the US National Library of Medicine (NLM) through annual version releases.  
Each new version may incorporate new vocabulary terms. For example, since 2018 a project is running in NLM to incorporate the vocabulary of the NCBI taxonomy in MeSH, starting with terminology Viruses and extending to Archaea, Bacteria, and Fungi.\footnote{\url{https://www.nlm.nih.gov/pubs/techbull/nd20/nd20\_mesh\_ncbi\_taxonomy.html}}
In addition, potential issues about inconsistencies, errors, or outdated information are also addressed during the annual maintenance of MeSH. For example, two closely related descriptors can be merged into one, as done with “Cascara” and “Rhamnus” in 2020.\footnote{\url{https://www.nlm.nih.gov/pubs/techbull/nd19/nd19\_medline\_data\_changes\_2020.html\#concept\_merge}} 
% Technically, this is usually done by removing one of the two descriptor identifiers from the new version of the vocabulary, and assigning the MeSH concept(s) of this deleted descriptor, as subordinate concept(s) in the remaining descriptor.
This continuous maintenance and evolution of MeSH ensures that this unique resource remains as up-to-date and free of errors as possible.

\begin{figure}     \includegraphics[width=0.47\textwidth]{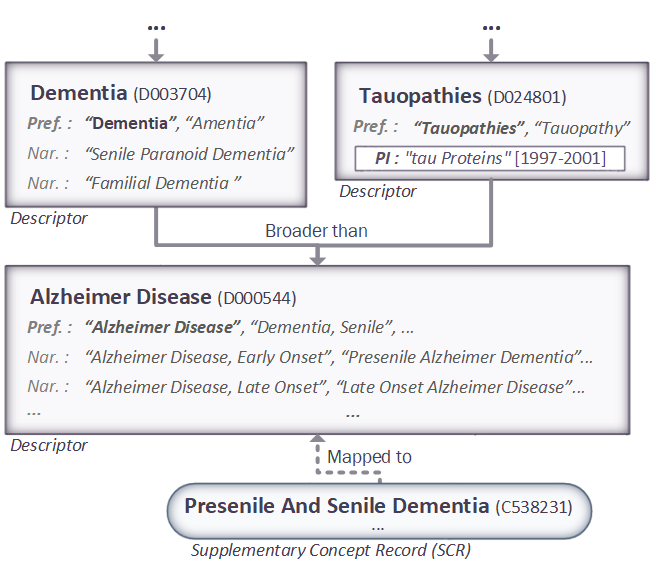}
\caption{MeSH concepts are grouped into descriptors, which are hierarchically organized and can also have a ``Previous Indexing'' note (PI). Each Supplementary Concept Record (SCR) is mapped to at least one descriptor. }
\label{fig:AD}       % Give a unique label
\end{figure}

In the MeSH hierarchy, each descriptor has exactly one \textit{preferred concept} and may also have some subordinate (narrower, broader, or related) concepts that attach additional terms to the descriptor. 
For example, the descriptor for Dementia (Fig.~\ref{fig:AD})
consists of three concepts. The preferred concept, which has two synonymous terms (``Dementia'' and ``Amentia'') and two narrower concepts with a single term each. 
The preferred concept is the reference point for defining the subordinate concepts as narrower, broader, or related. Therefore, we consider the preferred concept as the dominant entity representing the main topic of a descriptor.

% \begin{figure}   
% \centering
% \includegraphics[width=0.37\textwidth]{MeSHEvolution_AD_1.png}
% % figure caption is below the figure
% \caption{The internal structure of a MeSH descriptor comprising concepts and terms.}
% \label{fig:AD_1}       % Give a unique label
% \end{figure}

The MeSH descriptors are hierarchically organized so that most descriptors have at least one broader descriptor as a parent. For example, the ``Alzheimer Disease" (AD) descriptor has two parents, namely the descriptors ``Dementia'' and ``Tauopathies'', as shown in Fig.~\ref{fig:AD}.
Additionally, there are some top-level descriptors that have no parents and are the roots of the trees in the MeSH hierarchy, which are called MeSH \textit{trees}.\footnote{\url{https://www.nlm.nih.gov/mesh/intro_trees.html}}
% or just MeSH \textit{contexts}.
The MeSH trees are grouped into sixteen MeSH \textit{categories}\footnote{\url{https://www.nlm.nih.gov/bsd/disted/meshtutorial/meshtreestructures/index.html}}, and each descriptor belongs to one or more MeSH trees and corresponding MeSH categories. For example, the AD descriptor belongs to the ``Nervous System Diseases (C10)'' tree in the ``Diseases'' (C) MeSH category and to the ``Mental Disorders (F03)'' tree in the ``Psychiatry and Psychology'' (F).

The exact position of a descriptor in a tree is determined by one or more \textit{tree numbers} or \textit{tree paths}. 
Each MeSH tree number of a descriptor is extending a tree number of some parent, and recursively includes the tree numbers for a series of ancestors reaching up to the corresponding root.
For example, the AD descriptor has two tree numbers 
% in the ``Nervous System Diseases'' context (\url{C10.228.140.380.100},\url{C10.574.945.249}) and one in the ``Mental Disorders'' context (\url{F03.615.400.100}). Two of them are 
extending the tree numbers of ``Dementia'' (\url{F03.615.400.100}, \url{C10.228.140.380.100}) and one tree number extending the tree number of ``Tauopathies'' (\url{C10.574.945.249}).
% Though not technically forbidden, none of the tree numbers of a descriptor include any tree number of the descendants of this descriptor, as this would be against the purpose of organizing literature in increasing levels of specificity. 

The SCRs, that are also known as \textit{Supplementary Chemical Records}, have a similar conceptual structure to descriptors, with one preferred MeSH \textit{concept} and potentially some subordinate ones, but they are not part of any descriptor and are not directly included in the MeSH hierarchy. 
However, they are mapped to at least one descriptor. In Fig.~\ref{fig:AD} for example, the SCR ``Presenile And Senile Dementia''
is mapped to the AD descriptor as indicated by a dotted arrow towards the latter.
In practice, this means that when indexers in PubMed/MEDLINE use this SCR to annotate an article, the article will also get automatically annotated with the mapped descriptors~\cite{Nelson2001}. 
This mapping is important because it defines which descriptors cover the meaning of each SCR at the level of main MeSH topic annotations, that are primarily used for indexing and searching the literature.

\section{Related work}
\label{sec:RelatedWork}
The MeSH evolution is often studied in the broader context of the evolution of dynamic biomedical terminologies~\cite{DaSilveira2015}. 
In this area, the effort has often been on defining and studying elementary and composite changes, that require one or more basic operations, e.g. adding, removing, merging, splitting, editing, and moving elements of a biomedical terminology.
For instance, the CONCORDIA framework, which stands for ``CONcept and Change-Operation Representation for any DIAlect'' was proposed for representing, reporting, and documenting different types of change in medical terminologies~\cite{Oliver1999}, and MeSH was explicitly considered in this study. 
Studying different types of change in MeSH is also the focus of the work presented in this paper. The basic premise is that by studying the basic operations that lead to a change, one can identify the conceptual source of new elements. 

Though MeSH is not an ontology, it is often mentioned or even treated as such in the relevant literature, and it has also been transformed into a meta-ontology, in an effort to formally express all knowledge about semantically indexing MEDLINE~\cite{Abcckcr2005}.   
McCray and Lee~\cite{McCray2013} studied the evolution of MeSH in an ontological context, as a conceptualization of the biomedical domain.
They focused on the evolution of the MeSH category ``Psychiatry and Psychology'' (F), capturing and quantifying change at the level of descriptors, as well as at more fine-grained terminological and lexical levels.
% Their work is primarily related to the first of the research questions motivating this study, as 
In particular, they investigated whether change reflects the evolution of corresponding knowledge in the biomedical domain.
% revealing 
Their results reveal 
that change in MeSH reflects both the evolution of biomedical knowledge, as well as some internal ontological restructuring efforts, such as the separation of behaviors from disorders.

Recently, Balogh \textit{et al.}~\cite{Balogh2019} studied the evolution of MeSH as a network focusing on the addition and removal of links between MeSH descriptors, which they call ``attachment'' and ``detachment'' of links respectively. 
Interestingly, they investigated whether these re-wiring events are associated
with certain descriptor properties, such as the number of parents or descendants in the hierarchy. 
Their results suggest that old MeSH descriptors with many descendants appear to receive and lose children descriptors more than expected by chance.
% In particular, for descriptors that are new for the vocabulary, a preference is observed to having old descriptors with many children as their parents.
On the other hand, descriptors with many ancestors appear to receive and lose children descriptors less than expected by chance.

More recently, Cardoso \textit{et al.}~\cite{Cardoso2020} suggested the interlinking of distinct versions of MeSH developing a historical knowledge graph, to extend queries for biomedical literature retrieval and for supporting the maintenance of semantic annotations.
In particular, they introduce ``evolution connections'' between descriptor elements (e.g. terms) in different versions. 
In some cases, these evolutionary relationships can indicate the conceptual provenance for new descriptors, whereas in other cases they express more fine-grained internal restructuring, such as relocating a term from one MeSH concept to another.
Identifying the conceptual provenance of descriptors is also central in the work presented in this paper. 
However, the focus here is at the level of topics and the goal is to capture additional provenance connections, beyond MeSH concepts, namely through SCRs and Previous Indexing information.  

Some related studies also focus on the identification of the elements that need to change in MeSH and try to automate this process.
For example, Sari~\cite{Sari2020} proposed an approach for propagating changes already incorporated in the Gene Ontology into appropriate changes in MeSH. 
However, other studies attempt to predict the extension of MeSH based on different approaches.
For instance, Fabian \textit{et al.}~\cite{Fabian2012} proposed a method for finding siblings to a set of MeSH terms, analyzing the structure and the content of HTML pages on the Web.
Guo \textit{et al.}~\cite{YuWenGuo2014}, on the other hand, proposed a ``structure-based'' method for recommending new siblings for MeSH descriptors, that was exclusively based on the positions of existing terms in the MeSH hierarchy.
Eljasik-Swoboda \textit{et al.}~\cite{eljasik2019word} proposed an  embedding-based method for suggesting new sub-topics for existing topics of MeSH, combining both knowledge about the hierarchy and the analysis of documents already annotated with specific MeSH labels.

Other studies proposed approaches that beyond the analysis of annotated corpora and the structure of the hierarchy, they incorporate temporal information for the history of MeSH as well.
Tsatsaronis \textit{et al.}~\cite{Tsatsaronis2013} proposed a method to predict which MeSH descriptors should be expanded with new children. This method combined information about the number of articles annotated with each descriptor in PubMed with information about the hierarchical position of each descriptor in MeSH and temporal features that capture changes. 
Cardoso \textit{et. al}~\cite{Cardoso2018} also proposed a method for identifying  concepts that require revision, based on structural and temporal information, as well as information from other resources including the UMLS and article annotations in PubMed. Beyond the expansion of concepts with more children, their method also suggested other types of revision, such as removal and relocation.

Finally, the MeSH thesaurus has also been considered in some studies in the context of topic modeling and evolution in the biomedical domain. 
In these works, MeSH labels are treated as keywords to develop a network of keyword co-occurrence from a document corpus, where latent (meta-)topics can be detected as clusters or communities.
In this context, Castillo \textit{et al.}~\cite{castillo2016exploring} aligned such (meta-)topics of MeSH terms extracted from different time intervals based on the similarity of the corresponding sets. 
Then, they present an overview of the evolution of these matched (meta-)topics as a phylogeny-inspired network, where evolutionary events like merge and split can be identified.
Balili \textit{et al.}~\cite{Balili2020} propose the TermBall approach for both tracking and forecasting the evolution of such (meta-)topics of MeSH terms, treating them as evolving communities in the term co-occurrence network.
% In particular, they detect and predict six possible evolutionary events, namely emergence, growth, shrinkage, survival, merging, splitting, and dissolution of a (meta-)topic.
% Their results shed light in the evolution of the biomedical domain identifying the provenance of (meta-)topics detected in PubMed, but are not directly exploitable for semantic indexing.
The work presented here investigates evolutionary relations between biomedical topics as well, however, we focus on the level of MeSH descriptors, as used for indexing in PubMed, rather than the broader level of latent (meta-)topics reflecting the evolution of a domain.

% ~\cite{DaSilveira2015} States that biomedical terminologies often expand toward more fine-grained concepts.
In this study, we focus on newly added descriptors during the extension of MeSH. 
In particular, we study whether the meaning of each new descriptor has been covered by old descriptors previously, and if so, how its new position in the hierarchy relates to those of its predecessors. 
In contrast to most related work in the biomedical terminology evolution context, which focuses either on the operations to implement a change (addition, merge, etc) or on general features of the descriptors, such as depth in the hierarchy, we aim at characterizing the new descriptors according to their conceptual provenance. In other words, how they are related to their predecessors in previous versions of MeSH. In order to achieve this characterization, we investigate how we can identify the predecessors and introduce provenance types that provide new insight into the study of MeSH evolution.  

\section{A conceptual model for descriptor provenance}
\label{sec:ConceptualFramework}

In this section, we introduce a conceptual model to characterize and group new MeSH descriptors based on their conceptual provenance. 
That is, we investigate the cases of previous coverage of new descriptors during the extension of MeSH. 
We define the notion of Previous Host (PH), as a predecessor of a new descriptor, and describe categories of descriptors based on how these predecessors can be identified.
Subsequently, we introduce types of conceptual provenance, to characterise interesting cases of new descriptors, based on their current relation with each of their PHs in the hierarchy of MeSH.

\subsection{MeSH extension and provenance}
\label{sec:extensionProvenance}

\begin{figure}      \includegraphics[width=0.47\textwidth]{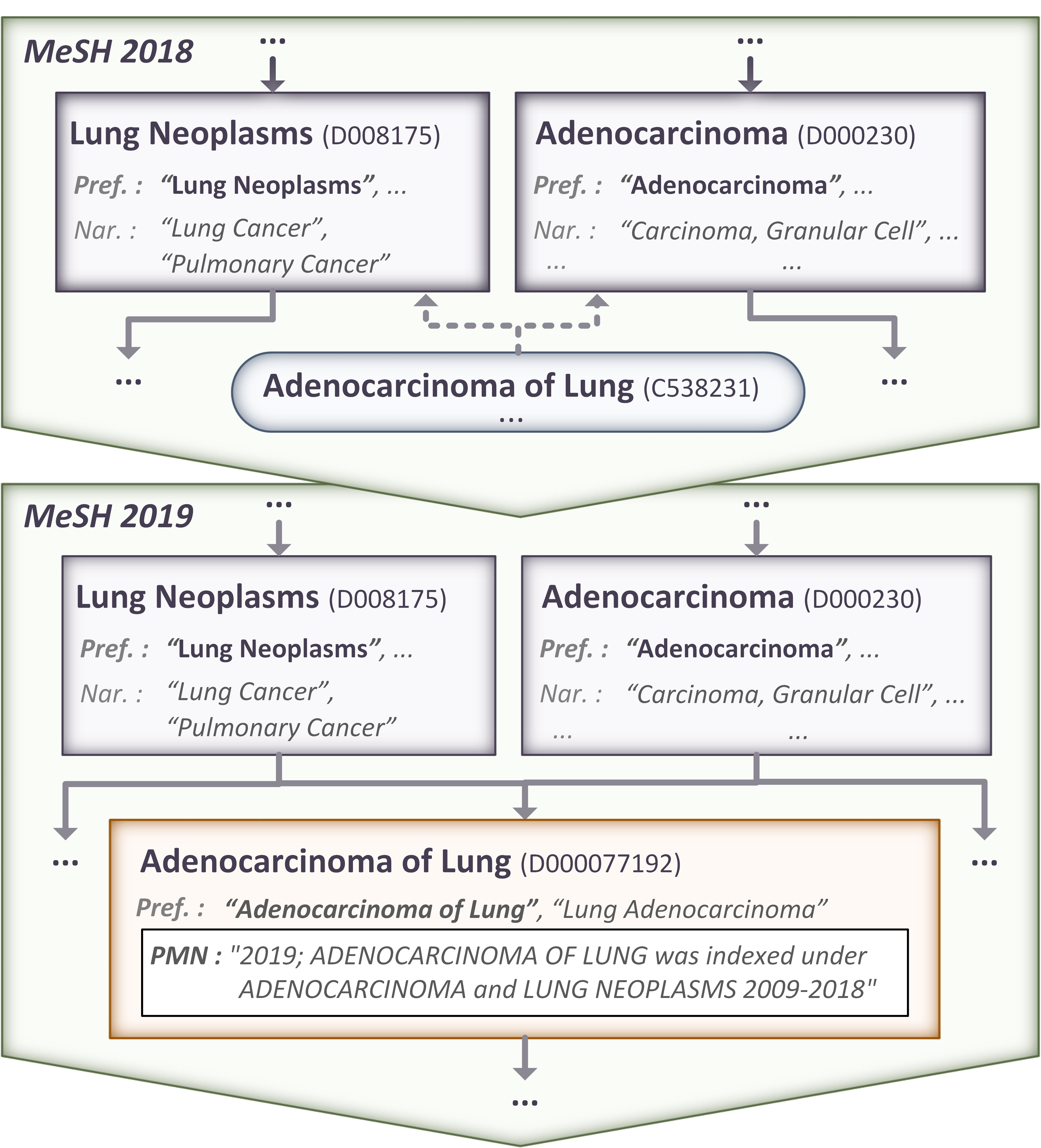}
\caption{The promotion of the SCR ``Adenocarcinoma of Lung'' into a descriptor in 2019.  }
\label{fig:Subdivision}       % Give a unique label
\end{figure}

As the MeSH hierarchy evolves, the new descriptors introduced may cover domain concepts that are not totally new to the vocabulary.
% most prior work is primarily on term level 
Some concepts may have been explicitly present in the previous version of MeSH.
In particular, a concept of a new descriptor may have been available as a subordinate concept of an old descriptor or as an SCR concept. 
The latter case, of turning an SCR concept into a descriptor, is usually reported in a textual note in the new descriptor, called \textit{Public MeSH Note} (PMN). For example, the ``Adenocarcinoma of Lung'' descriptor that was introduced in 2019, shown in Fig.~\ref{fig:Subdivision}, has a PMN field indicating its previous state as an SCR mapped to the ``Adenocarcinoma'' and ``Lung Neoplasms'' descriptors.   
% Since each concept may belong only to a single descriptor or SCR, it must be removed from any previous ones when added to the new descriptor. 
% As a rule, each concept is identified by the same unique identifier in different versions of MeSH.
Therefore, literature for ``Adenocarcinoma of Lung'' annotated in 2018, can be found with ``Adenocarcinoma'' and ``Lung Neoplasms'' topic labels.

In addition, even in cases where the concepts of the new descriptor have not been explicitly available as such, their meaning may have been implicitly covered by old descriptors.
Such information is usually available as a Previous-Indexing note (PI) in the new descriptors, as in the case of the ``Tauopathies'' descriptor in Fig.~\ref{fig:AD}. The PI note indicates that some old descriptors were used to annotate literature for the topic of the new descriptor, during a specific period prior to its introduction. 
For example, the ``tau Proteins'' descriptor was used to annotate articles about ``Tauopathies'' since 1997. This changed in 2002, with the introduction of a descriptor for ``Tauopathies''.  

% \todo[inline]{Motivate for the basic questions.}
In this work, we refer to the MeSH version of the introductory year of a new descriptor as \textit{version 1}, and the last year before \textit{version 1} as \textit{version 0}. 
In addition, we refer to such old descriptors that were used to annotate literature for the topic of the new descriptor in the version prior to its introduction (\textit{version 0}), as its \textit{Previous Hosts} (PHs). 
% For example, the PH for ``Tauopathies'' is ``tau Proteins''. That is, literature for the ``Tauopathies'' topic that has been annotated prior to 2001 can be found with ``tau Proteins'' annotations.
Apart from identifying the PHs of a new descriptor, the current relation of the new descriptor with its PHs is also important. 
For example, the new descriptor ``Adenocarcinoma of Lung'' was positioned in the hierarchy as a child of its two PHs (Fig.~\ref{fig:Subdivision}). Therefore, literature for ``Adenocarcinoma of Lung'' is still covered by ``Adenocarcinoma'' and ``Lung Neoplasms'' as done prior to the introduction of the new descriptor.
On the other hand, the new descriptor for ``Tauopathies'' is not hierarchically related to its PH ``tau Proteins''.
As a result, literature for ``Tauopathies'' is not covered by the ``tau Proteins'' descriptor after the introduction of the new descriptor in 2002. 
% This relationship of a descriptor with its PHs depends on the current version of MeSH, which we call \textit{reference version}, and can be \textit{version 0} or any later version.

Although the above cases are common, they are not the only types of relation one encounters between a new descriptor and its PH(s). 
Furthermore, as the MeSH hierarchy keeps evolving, the relation of a descriptor with one or more of its PHs can change in subsequent years, complicating the situation further.
Therefore, this relationship depends on the 
% current 
version of MeSH considered, which we call \textit{reference version}. 
In this work, aiming at a profound understanding and improved handling of new MeSH descriptors we investigate their origin. That is, whether they have been covered by descriptors (PHs) in the corresponding \textit{version 0}, and if so, what their current relation to each of these descriptors is. 
In order to better quantify and organize these cases, we define types of ``conceptual provenance'' for the new descriptors.
% and calculate their frequency of appearance during the last fifteen years. 

% ~~~~ define the PH
\subsection{Previous Hosts (PHs)}
A PH of a new descriptor is defined as a descriptor that was used to annotate articles for the topic of the new descriptor in the \textit{version 0} of the new descriptor. 
In that sense, we say that the PH used to cover the topic of the new descriptor for the purposes of literature annotation in \textit{version 0}. 
However, it is not required that a PH used to cover the topic exclusively.
That is, a PH may have been used for indexing other topics as well, apart from the topic of the new descriptor. Therefore, several new descriptors may share the same PH.
In addition, it is not required that a PH used to cover the topic of the new descriptor entirely. That is, a PH may have been used for indexing only part of a topic, for example in cases of a new high-level descriptor added to provide a new grouping of related topics.

A formal definition for a PH descriptor \textit{d0} for a new descriptor \textit{d1} can be based on the condition of \textit{topic-overlap} as follows: 
\begin{itemize}
    \item The \textbf{\textit{topic-overlap(d1, d0, v)}} is true when articles for the main topic of \textit{d1} used to be indexed under \textit{d0} in the MeSH version \textit{v}.
    In all other cases, \textit{topic-overlap(d1, d0, v)} is false.
    \item The \textbf{\textit{previous-host(d1, d0)}}, denoting that \textit{d0} is a PH of \textit{d1}, is true when \textit{topic-overlap(d1, d0, v)} is true, where \textit{v} is the \textit{version 0} of \textit{d1}.
    In all other cases, \textit{previous-host(d1, d0)} is false.
\end{itemize}

This definition of a PH focuses only in the MeSH version that precedes the introduction of the new descriptor (\textit{version 0}). 
Descriptors that used to cover the topic in older versions can be recursively described as the PHs of a PH and so on. 
However, our original motivation is to characterize each new descriptor based on whether its topic was already covered by MeSH, at the time of its introduction (\textit{version 1}), or not.
Therefore, in this work we do not track the history of each new topic in the distant past. 

As already discussed in subsection~\ref{sec:extensionProvenance} there are two types of coverage for a new descriptor in a previous version of MeSH. 
a) Explicit coverage, which is based on the conceptual structure of MeSH descriptors and SCRs into concepts, and 
b) implicit coverage, that can be identified based on the PI information. 
Based on the coverage type, we also characterize the corresponding PHs.
That is, an \textit{explicit} PH used to host a subordinate concept or used to be mapped from an SCR, that corresponds to the new descriptor.
On the other hand, an \textit{implicit} PH was used by the indexers for annotating articles that correspond to the topic of the new descriptor, without any explicit link with the latter in its conceptual structure.

Explicit PHs are of primary importance, as they provide strong conceptual links to the new descriptors. 
In our quest for a conceptual link to PHs, we focus on the preferred concept of each new descriptor. This is because the preferred concept is the dominant entity that represents the main meaning of a descriptor, as well as the vast majority of articles indexed with the descriptor. 
For descriptors that are new to the vocabulary, in the absence of any explicit PH, we exploit the PI field to identify any implicit PH.
The ``Tauopathies'' descriptor, shown in Fig.~\ref{fig:AD}, is such a case of a new descriptor without any explicit PH, where the PI field is exploited to identify the implicit PH ``tau Proteins''. 

\subsection{Provenance Categories}

For the purpose of identifying the PHs of a new descriptor \textit{d1}, we seek its preferred concept in the corresponding \textit{version 0} of MeSH. Based on whether and how we identify it in existing descriptors, we define four cases (\textit{categories}) of conceptual provenance, as depicted in Fig.~\ref{fig:Categories} and described below:
    
    \begin{figure}
    \includegraphics[width=0.47\textwidth]{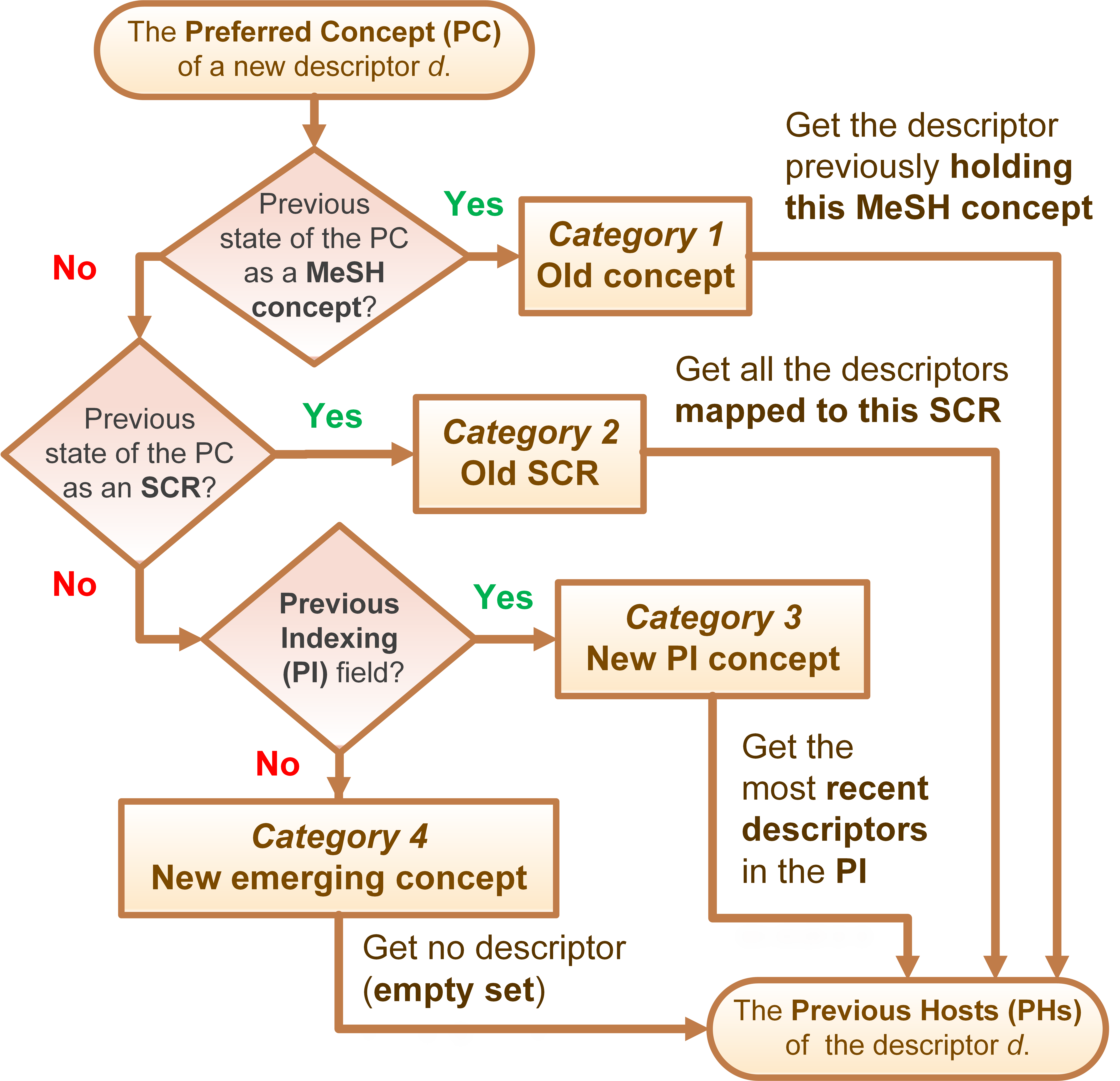}
    % \includegraphics[width=0.47\textwidth]{MeSHEvolution_Categories.png}
    \caption{Identifying the provenance category and the Previous Hosts (PHs) for a new descriptor. }
    \label{fig:Categories}       % Give a unique label
    \end{figure}
    
    \paragraph{Category 1. Old Concept:} 
    Although \textit{d1} is a new descriptor, its preferred concept is available in the previous version of MeSH (\textit{version 0}) as a subordinate concept of another descriptor \textit{d0}.
    In this case of explicit coverage, since \textit{d0} used to hold the preferred concept of \textit{d1}, \textit{topic-overlap(d1, d0, version 0)} is true. The descriptor \textit{d0} therefore, is an explicit PH of \textit{d1}.
    In addition, as each MeSH concept can only belong to a single descriptor in a given version of MeSH~\cite{Nelson2001}, \textit{d0} is the unique PH of \textit{d1}.
    
\begin{figure}
    \includegraphics[width=0.47\textwidth]{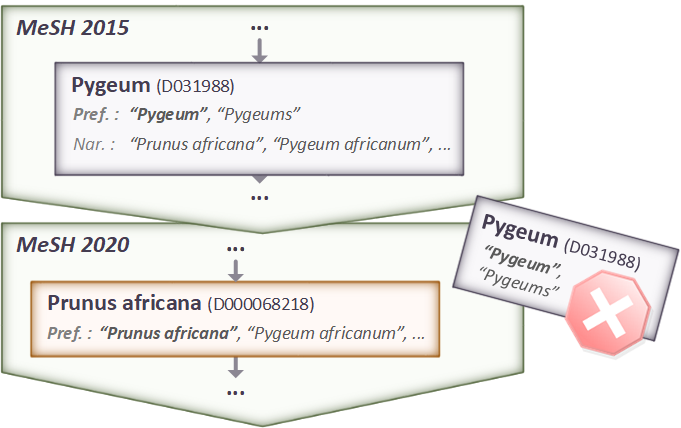}
    \caption{An example of descriptor \textit{succession}. }
    \label{fig:Succession}       % Give a unique label
    \end{figure}

    For example, ``Prunus africana'', shown in Fig.~\ref{fig:Succession}, introduced in 2016 as a descriptor, was a subordinate (narrower) concept of the ``Pygeum'' descriptor, which is not included in MeSH anymore. In this case, the unique PH of ``Prunus africana'' is the ``Pygeum'' descriptor, which explicitly included the concept ``Prunus africana'' in the version prior to the introduction of a dedicated descriptor for it.
    
    % For example, ``Pregnancy, Ovarian'' (D065172) introduced as a descriptor in 2015, was already available in MeSH 2014 as a subordinate concept in the ``Pregnancy, Ectopic'' (D011271) descriptor. 
    \paragraph{Category 2. Old SCR:} Alternatively, since the SCRs account for a large volume of domain concepts that are not included in MeSH descriptors~\cite{Bushman2015}, the preferred concept of \textit{d1} may have been available as a concept in an SCR \textit{scr}, prior to the introduction of \textit{d1} (\textit{version 0}). 
    In this second case of explicit coverage, for each  descriptor \textit{d0} mapped from \textit{scr} holds that
    % \textit{precedence(d1, d0)} is true. 
    % In addition, the mapping from \textit{scr}, which represents the preferred concept of \textit{d1}, to each \textit{d0} entails that 
    the literature indexed under \textit{scr} was also indexed under \textit{d0}. Therefore, \textit{topic-overlap(d1, d0, version 0)} is also true.
    As a result, each descriptor \textit{d0} is an explicit PH of \textit{d1}.
    For example, the descriptor ``Adenocarcinoma of Lung'' introduced in 2019, was previously available as an SCR mapped to the descriptors ``Lung Neoplasms'' and ``Adenocarcinoma'' (Fig.~\ref{fig:Subdivision}). These two descriptors are the explicit PHs of the new descriptor. 
    % For example, ``Synesthesia'' (D000080311) introduced as a descriptor in 2020, was already available in MeSH 2019 as SCR (C562460). 
    \paragraph{Category 3. New PI Concept:} 
    The preferred concept may be new, introduced together with the new descriptor \textit{d1}. 
    For such new descriptors, if previous-indexing (PI) information is available, this means that some other descriptors were previously used to index articles for the topic of \textit{d1} (new PI concept). 
    Therefore, the preferred concept of \textit{d1}, though new in the MeSH thesaurus, was previously indexed under some older descriptors with other concepts, hence implicitly covered by them.
    In such cases of implicit coverage, the PI descriptors that were used until the introduction of \textit{d1} are the ones with the most recent ending year in the accompanying period (\textit{version 0}). 
    Therefore, for each descriptor \textit{d0} that was used until the introduction of \textit{d1}, we have that
    % both \textit{precedence(d1, d0)} is true and
    \textit{topic-overlap(d1, d0, version 0)} is true. 
    As a result, the most recent PI descriptors are the implicit PHs of the new descriptor \textit{d1}. 

    For example, ``Zika Virus Infection'' 
    % (D000071243) 
    was introduced as a descriptor in 2015, and was not previously present as a concept in MeSH. 
    However, it is annotated with a PI note, revealing that the descriptors named ``Arbovirus Infections'' and ``Flavivirus Infections'' have been used for indexing literature relevant to ``Zika Virus Infection'' until 2015. 
    Therefore, these two descriptors are the PHs of ``Zika Virus Infection''.

    \paragraph{Category 4. New Emerging Concept:} 
    
    On the other hand, there exist new descriptors where no PI information is available, no PH can be identified and their PHs is an empty set. 
    Such totally new descriptors are expected to include emerging domain concepts without any significant presence in prior literature. 
    Therefore, the curators begin indexing articles for a domain topic previously not indexed under any specific MeSH descriptor. 
    
\begin{figure}      \includegraphics[width=0.47\textwidth]{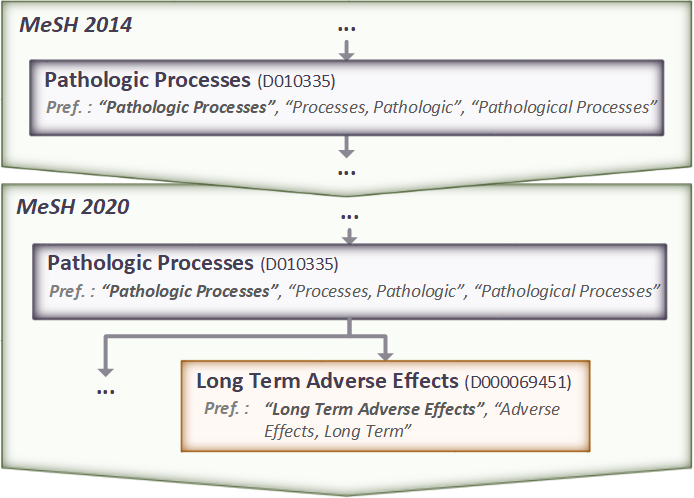}
    \caption{An example of descriptor \textit{emersion}. }
    \label{fig:Emersion}       % Give a unique label
    \end{figure}

    For example, ``Long Term Adverse Effects'' 
    % (D000069451) 
    was introduced in 2015 as presented in Fig.~\ref{fig:Emersion}.
    No ``Long Term Adverse Effects'' concept was previously present in MeSH and no PI information is available to report that articles for ``Long Term Adverse Effects'' were indexed under some particular descriptor until 2015. 
    Therefore, this totally new descriptor has no PHs at all.
    % However, the latter is annotated with a previous-indexing note, revealing that the descriptors named ``Arbovirus Infections'' and ``Flavivirus Infections'' have been used for indexing literature relevant to ``Zika Virus Infection'' until 2015. For ``Long Term Adverse Effects'' no such information is available.
    
\subsection{Provenance Types}
Having identified the PHs and the provenance category of each new descriptor, next we investigate the hierarchical relation of the new descriptor with each one of its PHs. 
This relation starts with the introduction of the new descriptor in the MeSH hierarchy but may change in the course of the years, as the hierarchy evolves further. 
Therefore, to characterise the relation of a new descriptor \textit{d1} with a PH \textit{d0} in the context of a given \textit{reference version} of MeSH, we focus on two basic properties of this relation in the corresponding hierarchy. 
Namely the \textit{relation type} of \textit{d1} with \textit{d0} and the \textit{distance} between them in the hierarchy.
\begin{itemize}
    \item The \textbf{\textit{relation\_type(d1, d0)}} is: (a) \textit{ancestor} when \textit{d1} has at least one \textit{tree number} that includes a tree number of \textit{d0}, (b) \textit{descendant} when \textit{d0} has at least one \textit{tree number} that includes a tree number of \textit{d1}, (c) \textit{unrelated} when none of the \textit{tree numbers} of \textit{d0} includes or is included in any of the  \textit{tree numbers} of \textit{d1}, and (d) \textit{undefined} when \textit{d0} is not present in the \textit{reference version} of MeSH.\footnote{Although the four relation types are expected to be mutually exclusive as a rule, the types (a) \textit{ancestor} and (b) \textit{descendant} can coexist in exceptional cases.  
    In particular, analyzing the last fifteen versions of MeSH we found six such descriptor pairs where one descriptor is both an ancestor and a descendant of the other descriptor. All these pairs were introduced prior to 2014 and none of them corresponds to a descriptor and its PH.}
    % If the \textit{d0} is not present in the given version of the hierarchy the \textit{relation\_type(d1, d0)} is \textit{undefined}.
    % If the d0 is \textit{null} the inclusion(d1, d0) is also null.
    \item The \textbf{\textit{distance(d1, d0)}} is the number of other descriptors included in the shortest path connecting \textit{d1} and \textit{d0}. 
    % A path connecting two descriptors is identified considering the overlap of any \textit{tree number} of the one descriptor to a \textit{tree number} of the other.
    A path connecting two descriptors comes from any pair of overlapping \textit{tree numbers} of the two descriptors.  
    Two \textit{tree numbers} are overlapping if one of them includes the other, or if both include the same \textit{tree number}, which corresponds to a common ancestor.
    If no such pair of \textit{tree numbers} exists for \textit{d0} and \textit{d1}, then they are located in positions of the hierarchy that are not connected. In this case, the \textit{distance(d1, d0)} can be considered to be infinite.
    If \textit{d0} is not present in the \textit{reference version} of the hierarchy the \textit{distance(d1, d0)} is \textit{undefined}.
    % \item The \textbf{\textit{distance(d1, d0)}} is the number of other descriptors included in the shortest path connecting \textit{d1} and \textit{d0}. 
    % If \textit{d0} is located in a position in the hierarchy that is not connected with \textit{d1}, then the \textit{distance(d1, d0)} can be considered to be infinite.
    % If the \textit{d0} is not present in the \textit{reference version} of the hierarchy the \textit{distance(d1, d0)} is \textit{undefined}.
    % If the d0 is \textit{null} the distance(d1, d0) is also null.

    % For example the distance( ``Adenocarcinoma of Lung'', ``Lung Neoplasms'') is zero (see Fig.~\ref{fig:Subdivision}), whereas the distance( ``Prunus africana'', ``Pygeum'') is infinite (see Fig.~\ref{fig:Succession}).
\end{itemize}

For example, the relation between the new descriptor ``Adenocarcinoma of Lung'' and its PH ``Lung Neoplasms'' has \textit{ancestor} \textit{relation\_type} and \textit{zero} \textit{distance} with MeSH 2019 as \textit{reference version} (see Fig.~\ref{fig:Subdivision}). 
On the other hand, the current relation of ``Prunus africana'' and its PH ``Pygeum'', in the context of MeSH 2020 \textit{reference version}, has \textit{undefined} \textit{relation\_type} and \textit{undefined} \textit{distance} (see Fig.~\ref{fig:Succession}).
% For example, the inclusion( ``Adenocarcinoma of Lung'', ``Lung Neoplasms'') is ancestor and the distance( ``Adenocarcinoma of Lung'', ``Lung Neoplasms'') is zero in the context of MeSH 2019 (see Fig.~\ref{fig:Subdivision}), whereas the inclusion( ``Prunus africana'', ``Pygeum'') is undefined and the distance( ``Prunus africana'', ``Pygeum'') infinite (see Fig.~\ref{fig:Succession}).
Based on the \textit{relation\_type} and the \textit{distance} of the relation between a new descriptor \textit{d1} and a PH \textit{d0} we also define some cases of interest, which we call conceptual provenance \textit{types}. 

% As the motivation of this study is to facilitate the current grouping and handling of descriptors in applications such as automated semantic indexing, we focus on a current situation, i.e. the most recent version of the hierarchy (MeSH2020). However, the 

% In this study we focus on the current situation, i.e. the most recent version of the hierarchy (MeSH2020). We compare this to the point where the change has happened, but we do not follow the detailed history of changes. Hence, regarding the relation of a new descriptor with its PHs, we define the following cases of interest, which we call conceptual provenance \textit{types}:

    \paragraph{Type 0. Emersion: No PH found.}
    For new descriptors in category 4, where no PH can be identified. 
    In these cases there is no PH for which to investigate the current relation, therefore we define the trivial type of provenance \textit{emersion}, which includes all descriptors of provenance \textit{category 4} and only descriptors of \textit{category 4}.
    
    This exceptional type of provenance does not reflect the relationship with any PH, therefore it is not based on \textit{relation\_type} and \textit{distance} in a specific \textit{reference version} of MeSH.
    The meaning of such a completely new descriptor is emerging when the new descriptor is introduced and is characterized as emergent hereafter. 
    The “Long Term Adverse Effects” descriptor introduced in 2015, is an example of \textit{emersion} (see Fig.~\ref{fig:Emersion}).

    \paragraph{Type 1. Succession: relation\_type(d1, d0) = undefined and distance(d1, d0) = undefined.}
    For some new descriptors a PH can be no longer present in the \textit{reference version} of MeSH. In this case, \textit{d1} is considered one of the successors of \textit{d0}, because at least some of the articles that used to be annotated with \textit{d0}, in \textit{version 0} for \textit{d1}, are annotated with \textit{d1} instead, in the \textit{reference version} of MeSH. 
    In the example of Fig.~\ref{fig:Succession}, the new descriptor “Prunus africana” is a case of succession, as its PH is not available in the context of the \textit{reference version}, MeSH 2020.
    % (D000068218), introduced in 2016 as a descriptor, was a subordinate (narrower) concept of the “Pygeum” (D031988) descriptor, which was removed in 2016 

\begin{figure}      \includegraphics[width=0.47\textwidth]{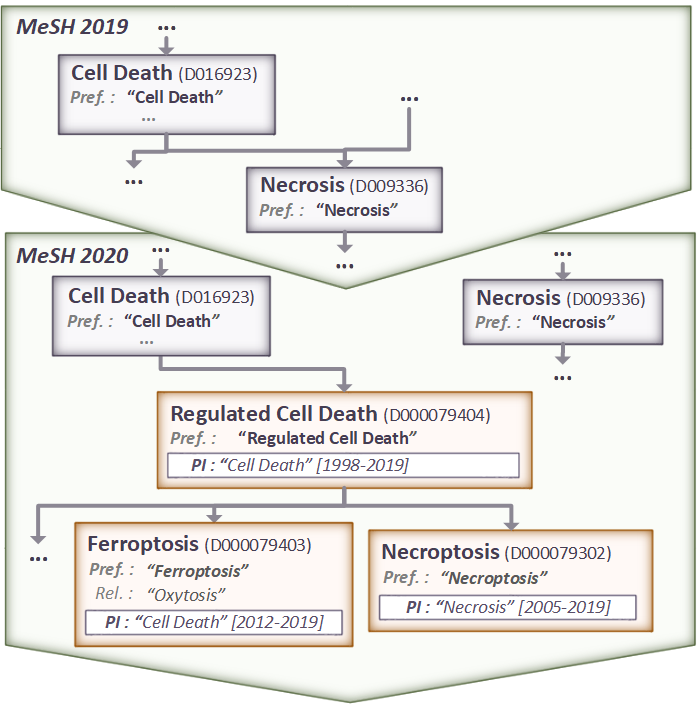}
    \caption{``Regulated Cell Death'' as a \textit{subdivision} of ``Cell Death'', the \textit{submersion} of ``Ferroptosis'' and the \textit{detachment} of ``Necroptosis''. }
    \label{fig:Submersion}       % Give a unique label
    \end{figure}

    \paragraph{Type 2. Subdivision: relation\_type(d1, d0) = ancestor and distance(d1, d0) = 0.}
    A new descriptor \textit{d1}, whose PH \textit{d0} has become its parent. In this case, \textit{d0} covers the topic of the new descriptor entirely, but \textit{d1} supports the partition of the corresponding literature into more fine-grained conceptual sets. 
    This is the most expected type of relation between new descriptors and their PHs, as the vocabulary evolves towards more detailed descriptors to support more precise topic annotations.
    % In the example of Fig.~\ref{fig:Subdivision}, the ``Adenocarcinoma of Lung'', introduced in 2019, used to be an SCR mapped to ``Lung Neoplasms'' and ``Adenocarcinoma'', both of which became its parents. 
    % Similarly, ``Regulated Cell Death'' depicted in 
    In the \textit{subdivision} example of Fig.~\ref{fig:Submersion}, ``Regulated Cell Death'' introduced in 2020, used to be indexed under ``Cell Death'' until 2019, which became its parent.
    % Similarly, “Hepatic Infarction” (D000081011), also introduced in 2019 as a descriptor with a new concept, was previously indexed under “Liver Diseases” (D008107) and “Infarction” (D007238), which are its current parents.
        
    \paragraph{Type 3. Submersion: relation\_type(d1, d0) = ancestor and distance(d1, d0) $>$ 0.}
    
    A new descriptor \textit{d1}, whose PH \textit{d0} has become an ancestor, but not a parent. This is similar to \textit{subdivision}, as they both are characterized by \textit{ancestor} \textit{relation\_type}, but at least one other descriptor appears between \textit{d0} and \textit{d1} in the hierarchy. 
    This is also in accordance with the evolution towards more detailed descriptors, as the \textit{d0} keeps covering the topic of the new descriptor entirely. 
    % This is also consistent with the situation prior to the introduction of the new descriptor, as the PH keeps covering the topic of the new descriptor. 
    However, the distance between them suggests that intermediate levels of detail are also available.
    
    ``Ferroptosis'', introduced in 2020 (Fig.~\ref{fig:Submersion}), is an example of \textit{submersion}, as it was indexed under ``Cell Death'' until 2019, which is now an ancestor but not a parent of it. In this example, ``Regulated Cell Death'' which acts as the intermediate level of detail, was also introduced together with ``Ferroptosis'', which explains why ``Ferroptosis'' articles were previously indexed under ``Cell Death'' instead of ``Regulated Cell Death''.
    
    \paragraph{Type 4. Overtopping: relation\_type(d1, d0) = descendant.}
    
\begin{figure}      \includegraphics[width=0.47\textwidth]{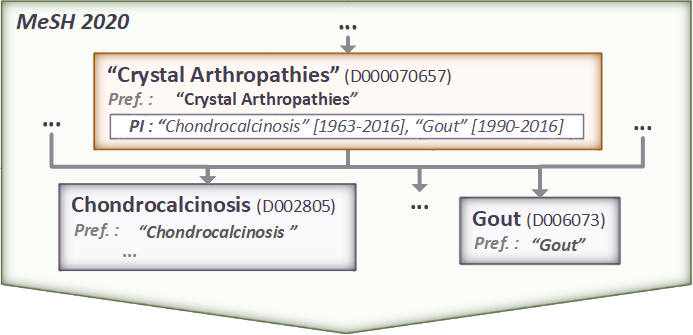}
    \caption{The ``Crystal Arthropathies'' \textit{overtopping} its PHs. }
    \label{fig:Overtopping}       % Give a unique label
    \end{figure}

    A new descriptor \textit{d1}, whose PH \textit{d0} has become its descendant. 
    In this case, although literature for the new topic used to be indexed under \textit{d0} in the past (\textit{version 0}), \textit{d1} is an ancestor of \textit{d0} in the \textit{reference version} of MeSH, hence broader than it. 
    % In this case, the new descriptor is semantically broader than the corresponding PH. 
    Such new descriptors provide a new grouping of the old topics, potentially enhanced with additional terms for the aggregate topic. 
    In the example depicted in Fig.~\ref{fig:Overtopping}, ``Crystal Arthropathies'', introduced in 2017, has two implicit PHs, as it was indexed as ``Chondrocalcinosis'' and ``Gout'' until 2016. Both of them are children of ``Crystal Arthropathies'' in 2020, hence overtopped by it.
    
    Such cases seem less expected than the ones with \textit{ancestor} \textit{relation\_type} (\textit{subdivision} and \textit{submersion}), as this situation suggests that \textit{d0} used to cover only a part of the topic of \textit{d1}. 
    In addition, \textit{overtopping} is less interesting from a practical point of view, as the use of the narrower descriptor that covers a topic is a common MeSH-indexing practice~\cite{Nelson2001}. 
    % This case seems quite unexpected, as the situation expressed is inconsistent with the one prior to the introduction of the new descriptor. 
    Therefore, though different levels of detail may exists between the new descriptor and its descended PH, splitting this small group of cases based on the \textit{distance} would not add any particular value. 
    
    \paragraph{Type 5. Detachment: relation\_type(d1, d0) = unrelated.}
    
    A new descriptor \textit{d1} that is not related to its PH \textit{d0} with any of the above relations. In this case, \textit{d1} is detached from \textit{d0}, placed in a position without the one being included by the other. 
    In the example of Fig.~\ref{fig:Submersion}, ``Necroptosis'', introduced in 2020 as a child to ``Regulated Cell Death'' in the ``Phenomena and Processes'' MeSH category (G), was previously indexed as ``Necrosis''. Although ``Necrosis'' used to be a child of ``Cell Death'' in 2019, in 2020 it belongs only to the ``Diseases'' MeSH category (C) and is not directly related to ``Necroptosis''. Therefore, we consider the ``Necroptosis'' descriptor to be detached from its PH ``Necrosis'' in 2020. 
    
    \begin{table*}
\caption{Provenance codes characterizing the relationship of a new descriptor with a PH, encoding categories and types as prefixes and suffices respectively. The exceptional case of \textit{emersion} type corresponds to code 4.0.}
\label{tab:codes}       % Give a unique label
\centering
\begin{tabular}{M{0.03\linewidth}L{0.12\linewidth}L{0.13\linewidth}M{0.07\linewidth}M{0.11\linewidth}M{0.11\linewidth}M{0.12\linewidth}}
\multicolumn{2}{c}{\textbf{Provenance Type}} &\multicolumn{2}{c}{\textbf{Properties}}& \multicolumn{3}{c}{\textbf{Provenance Category}}        \\
\multicolumn{2}{c}{}    & \textbf{relation\_type} & \textbf{distance} & \textbf{1. Old concept} & \textbf{2. Old SCR} & \textbf{3. New PI concept} \\
\textbf{.1}   & \textbf{Succession}  & undefined & undefined &	 1.1         & 2.1     & 3.1         \\
\textbf{.2}   & \textbf{Subdivision} & ancestor & 0 &	 1.2         & 2.2     & 3.2         \\
\textbf{.3}   & \textbf{Submersion}  & ancestor & $> 0$ &	 1.3         & 2.3     & 3.3         \\
\textbf{.4}   & \textbf{Overtopping} & descendant & $\geq 0$ &	 1.4         & 2.4     & 3.4         \\
\textbf{.5}   & \textbf{Detachment}  & unrelated & $\geq 1$ &	 1.5         & 2.5     & 3.5         
\end{tabular}
\end{table*}
    
     \begin{figure}      \includegraphics[width=0.47\textwidth]{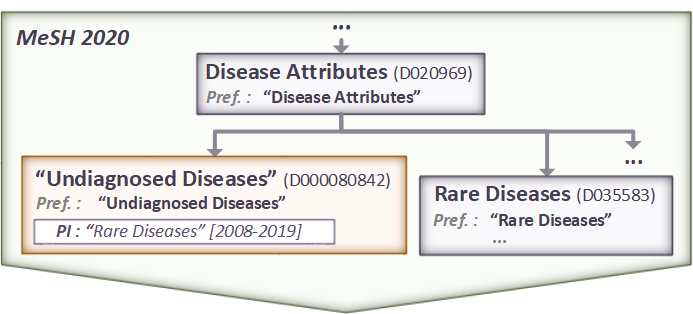}
    \caption{The \textit{detachment} of ``Undiagnosed Diseases'' from ``Rare diseases''. }
    \label{fig:Parity}       % Give a unique label
    \end{figure}   
    
    Detached descriptors may be positioned quite close to their PH in terms of \textit{distance}, but are not related as ancestors or descendants to it. 
    In the example of Fig.~\ref{fig:Parity}, “Undiagnosed Diseases”, introduced in 2020 as a child descriptor to “Disease Attributes”, was previously indexed under “Rare Diseases” which is also a child of “Disease Attributes”. 
    However, we consider that “Undiagnosed Diseases” is detached from its PH “Rare Diseases”, as their topics are effectively disjoint. That is, neither of the two topics includes the other in the \textit{reference version} (MeSH 2020).

    % For example, “Drug Collateral Sensitivity”, introduced in 2020 as a child to “Drug Interactions”, was previously indexed under “Drug Resistance” (Fig.~\ref{fig:Detachment}). In this case, the new parent “Drug Interactions” is a sibling to the PH “Drug Resistance”, as they are both children of “Pharmacological Phenomena”. However, we consider that “Drug Collateral Sensitivity” is detached from its PH “Drug Resistance”, as their topics are effectively disjoint. That is, neither of the two topics includes the other in the given version of MeSH.
    % \begin{figure}      \includegraphics[width=0.47\textwidth]{MeSHEvolution_6_Detachment.png}
    % % figure caption is below the figure
    % \caption{The detachment of the ``Drug Collateral Sensitivity'' descriptor. }
    % \label{fig:Detachment}       % Give a unique label
    % \end{figure}
    
\paragraph{Provenance codes:}

In order to easily refer to both category and type of conceptual provenance, we adopt a composite \textit{provenance code}, with a prefix indicating the category of a descriptor and a suffix indicating the \textit{type} of its relation to some PH, separated by a dot, as shown in Table~\ref{tab:codes}. 
For example, the \textit{provenance code} for ``Necroptosis'' (Fig.~\ref{fig:Submersion}) is 3.5 indicating a provenance \textit{category 3} for ``new PI concept'', as the PH has been identified based on PI information, and a provenance type 5 for \textit{detachment} from ``Necrosis''. 
Similarly, the provenance code for ``Prunus africana'' (Fig.~\ref{fig:Succession}) is 1.1 with \textit{category 1} for ``old concept'', and type 1 for \textit{succession} of ``Pygeum''.
In the special case of type 0, \textit{emersion}, the preferred concept of the new descriptor is by definition new (\textit{category 4}), hence, all \textit{emersion} cases have the trivial provenance code 4.0.

\begin{figure}      \includegraphics[width=0.47\textwidth]{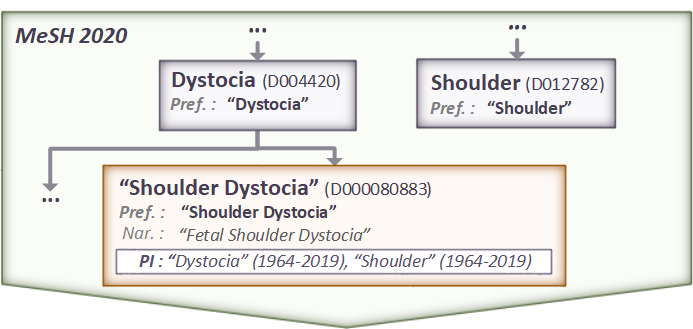}
\caption{The detachment of the ``Shoulder Dystocia'' descriptor has two provenance codes, namely, code 3.2 for the subdivision of the PH ``Dystocia'' and code 3.5 for the detachment from the PH ``Shoulder''. }
\label{fig:MulitpleCodes}       % Give a unique label
\end{figure}

As some new descriptors can have more than one PH, the provenance types described above are not mutually exclusive.\footnote{In addition, multiple provenance types could also be the result of exceptional cases, where a new descriptor is related to a single PH, both as an ancestor and as a descendant.} 
Therefore, a new descriptor can have multiple \textit{provenance codes}. 
This is not true for the provenance categories, therefore all \textit{provenance codes} of a specific new descriptor begin with the same prefix.
For example, ``Shoulder Dystocia'' depicted in Fig.~\ref{fig:MulitpleCodes}, was introduced in 2020 as a child descriptor to “Dystocia”. Articles for shoulder dystocia were indexed as both ``Dystocia'' and ``Shoulder'' until 2019, hence it is both a case of \textit{subdivision} (3.2) of the PH ``Dystocia'' which became its parent and a case of \textit{detachment} (3.5) from the PH ``Shoulder'' which is not directly related with the new descriptor.
% the exception of 3.0
% Also no overlap for 1.x codes

% \todo[inline]{cases of multiple types for the same PH - i.e. the definition of detachment as "none of the above"}

\section{Computing provenance of MeSH descriptors}
\label{sec:DataAnalysis}

\begin{figure*}  
\center
\includegraphics[width=0.98\textwidth]{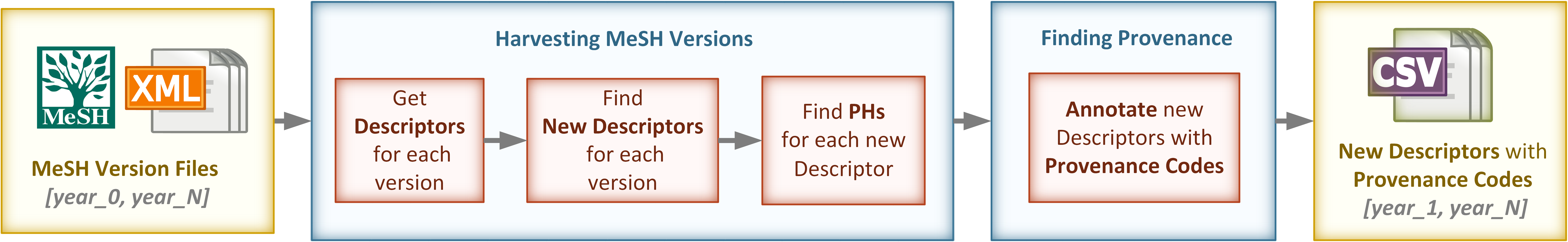}
\caption{The computational process for identifying new descriptors and annotating them with provenance codes. }
\label{fig:Schema}       % Give a unique label
\end{figure*}

In this section, we describe the computational tools developed for the automated identification of new MeSH descriptors, their PHs, and \textit{provenance codes}, in the context of the conceptual model introduced in section~\ref{sec:ConceptualFramework}.
These tools, access the original source files of MeSH\footnote{\url{https://www.nlm.nih.gov/databases/download/mesh.html}}, as provided by NLM, in the MeSH XML format\footnote{\url{https://www.nlm.nih.gov/mesh/xmlmesh.html}}. Therefore, all available information is accessible by the tools and any new versions of the hierarchy can be directly incorporated upon release. 
Figure~\ref{fig:Schema} illustrates the sequence of processing steps that are involved in relating new descriptors to their PHs.
The source code of the tools is openly available on GitHub.\footnote{\url{https://github.com/tasosnent/MeSH_Extension}}

% \begin{figure}      \includegraphics[width=0.47\textwidth]{Schema.png}
% % figure caption is below the figure
% \caption{The steps of the computational process developed for identifying new descriptors and annotating them with provenance codes. }
% \label{fig:Schema}       % Give a unique label
% \end{figure}

\subsection{Harvesting MeSH versions}
As mentioned in previous sections, we focus our analysis on the provenance of descriptors that are present in a \textit{reference version} of MeSH, namely the latest one. Therefore, we do not process descriptors that appear and disappear in various older versions. However, we are still interested in annotating descriptors that appear in older versions and remain available in the \textit{reference version}. As a result, we need to process older versions as well, covering a period from \textit{year\_0} to \textit{year\_N}.    

In particular, the process starts with the harvesting of MeSH files for different versions of the hierarchy. 
This step begins with parsing the basic XML file for each year
% (\textit{YYYY}), conventionally named \textit{descYYYY.xml}, 
to extract the descriptors available in this version.
This set of descriptors is then compared to those of the previous year to identify the new ones. 
% To validate the correctness of the identified new descriptors per year, we compared the results for random years with the information available in corresponding NLM reports for these years. All deviations in new descriptor identification were explained by the fact that the reports are not retrospectively re-considered, as done in our work. 
The same process is repeated for each year, with the exception of the very first one, for which no previous version is available.
Apart from the basic file comprising the MeSH descriptors, the XML file of the SCRs is also parsed for each version, to extract the corresponding set of available SCRs. These are needed for the identification of provenance categories and types.
% (\textit{suppYYYY.xml}). 

% Extracting descriptor attributes
\paragraph{Extracting descriptor attributes:}
 For the descriptors of interest, a number of attributes need to be extracted, in order to help us trace its provenance. The most important attribute is the MeSH code of the descriptor, which is the unique identifier considered for checking descriptor existence and identity. Other relevant information includes the positions of a descriptor in the hierarchy (\textit{tree numbers}), its preferred concept, and the content of the PMN and the PI fields. 
Most parts of this attribute extraction step are quite straightforward, as we primarily rely on the unique identifiers of the entities involved in the analysis. 
For example, the information needed for identifying the earlier status of a descriptor as a subordinate concept in its \textit{version 0}, is the unique concept identifier of its preferred concept. This is because we need to compare this identifier with the identifiers of subordinate concepts of any descriptor in \textit{version 0}.  

However, automated extraction of information from the PMN and PI fields proved more challenging as these fields contain information in semi-structured text, meant to be read by humans. Therefore the structure of this text is inconsistent, while descriptors and SCRs are mentioned with their current preferred terms, instead of the corresponding unique identifiers. 
Consequently, we adopted a semi-automated approach, based on regular expressions, in order to extract information from these fields. In the large majority of cases, we managed to minimize the required manual effort as described below. 

% Extraction from the PMN field
\paragraph{Extraction from the PMN field:}
The PMN (\textit{Public MeSH Note}) field of a MeSH descriptor typically consists of sentences separated by semicolons and may provide varying information, such as the year the descriptor was introduced and changes in the preferred term. 
 Of particular interest for this work, are PMN sentences that report the earlier status of the descriptor as an SCR. This is done with expressions of the form ``\textit{X was indexed under Y}'', where \textit{X} is the SCR and Y comprises one or more descriptors together with the corresponding time periods, as shown in the example of Fig.~\ref{fig:Subdivision}. This is useful as in some cases an SCR that gets ``promoted'' to descriptor may undergo some minor term modifications and receive a new identifier. In such cases, exploiting the PMN is the only way to identify the old SCR for the new descriptor, which would otherwise be considered totally new. 

Therefore, when attempting to associate a new descriptor to an earlier SCR, we start by comparing the identifier of the preferred concept of the descriptor to the concept identifiers in earlier SCRs. If this exact-match search fails, we resort to the use of the PMN expressions mentioned above. In particular, we first use regular expressions to extract from the PMN field the preferred term (\textit{X}) of the old SCR and map it to some SCR identifier in the corresponding version of MeSH. 
% 404 / 550 = 74%
In our analysis, this method managed to automatically identify the missing links for the majority of cases (74\%) where the PMN field matches the ``\textit{X was indexed under Y}'' expression and the exact-match search fails. 
% This percentage corresponds to about (X\%) of the total cases where a PMN field was available and the exact match failed.

For the few remaining cases, we calculated the similarity of \textit{X} and the current descriptor name to earlier SCR terms. Based on this similarity, the system produced best-match suggestions, which were confirmed manually. More details about this method are available in a technical report available online~\cite{nentidis2021all}.
There is also a small number of cases where more than one old SCRs is reported in the PMN field. In such cases, only the first SCR was considered, as this usually corresponds to the preferred concept of the new descriptor, representing its central meaning. 
% Manual comparison of the PHs found based on the SCR extracted by the pattern-based method with the ones mentioned in \textit{Y} revealed an agreement of about 93\%.  with disagreements often attributed to inconsistencies and typos in the PMN field.
% ~\footnote{For example, D059847 with PMN: ``2012; HLA-DRB5 was indexed under 1992-2011'' }.  

% Extraction from the PI field
\paragraph{Extraction from the PI field:}
The PI (\textit{Previous-Indexing}) field of a MeSH descriptor can be used to link a new descriptor to old ones when such a link is not provided explicitly, that is by a previous state of the descriptor as a subordinate concept or SCR. The PI field contains a list of semi-structured notes in English. 
Each note usually consists of the relevant descriptors for a previous period, often followed by the corresponding time period in parentheses (Fig.~\ref{fig:AD}).
Exploiting this pattern we used regular expressions to extract the terms and the corresponding time periods.\footnote{Some exceptions not fitting the patterns were identified and handled manually.}
% may give small inflation to emersion cases, by failing to extract some previous hosts. 
In cases where the PI field consists of multiple notes, all the descriptors with the most recent end year are considered as PHs, as done for ``Shoulder Dystocia'' in the example of Fig.~\ref{fig:MulitpleCodes}. Any older PI elements are neglected.

% Selecting provenance type
\paragraph{Selecting provenance type:}
In the last part of the MeSH harvesting step, each new descriptor is annotated with conceptual provenance codes. 
In particular, the first step is to select the provenance category based on the previous state of the current preferred concept as a subordinate concept or an SCR concept in the corresponding \textit{version 0}, as depicted in the schema of Fig.~\ref{fig:Categories}.
Then, the provenance type is selected, based on the current relation of the new descriptor to its PHs, which have been identified by the extraction process. 
Combining the provenance types with the category, the complete set of provenance codes is formed. 
The end result is a collection of all the new descriptors that have been introduced during the period considered
% ~\footnote{Between \textit{year\_0} and \textit{year\_N}}
and remain available in the \textit{reference version} of MeSH. These descriptors are annotated with their basic information and provenance annotations and stored in CSV files named after the year that corresponds to the $vesrion_1$ for each descriptor.  

% In the second step of the MeSH-data-analysis process, the distinct CSV files with the new descriptors and provenance annotations are parsed and analyzed for the production of alternative and complementary overall diagrams.
% In brief, the produced diagrams present the frequencies of provenance categories, types and codes per year or in total, as well as the correlation among them or with other properties of the descriptors. For correlation in particular, some pair plots~\footnote{\url{https://seaborn.pydata.org/generated/seaborn.pairplot}} are produced considering the provenance types as binary variables. Each pair plot combines a scatter plot and a linear model for each pair of types, with a bar plot depicting the univariate distribution of the data for each type.

\section{Analysis of MeSH versions}
\label{sec:results}

\subsection{Analytical setting}
\label{ssec:analytical_setting}

In this work, we applied the process presented in section~\ref{sec:DataAnalysis} on the source files of all versions of MeSH published in the last fifteen years. In this manner, we identified and annotated all the descriptors introduced during this period, considering MeSH 2020 as the \textit{reference version}. In other words, we are interested in the current status of the descriptors, but we use the year of their introduction \textit{version 1}, in order to identify their previous hosts (PHs) and provenance category.
The result of the computational processing is a CSV file for each MeSH version, comprising the new descriptors introduced this year and their provenance annotations. 

As a final step, these files\footnote{ \url{https://raw.githubusercontent.com/tasosnent/MeSH_Extension/main/NewDescriptors_2006_2020.csv}} are parsed and analysed to produce statistics and diagrams that provide alternative views of conceptual provenance in the course of MeSH expansion in order to answer the basic questions driving this study.
In particular, the diagrams that are generated present the frequencies of provenance categories, types, and codes per year of introduction and in total.
% , as well as the correlation among categories, types, codes, and other attributes of the descriptors. 
Based on these diagrams, we attempt to answer the basic questions driving this study and identify patterns and observations that may be of interest for understanding the dynamics of the extension of MeSH.
% that may be of interest in MeSH development, as well as further use of data annotated with MeSH descriptors.

\subsection{Overview of new descriptors and their provenance}
\label{sec:initial_numbers}

Table~\ref{tab:distribution} presents the distribution of new descriptors into
provenance categories and types. 
In total, 6,915 descriptors were introduced in MeSH since 2006 and were retained until 2020. This corresponds to an extension of about 30\%, compared with the 22,997 descriptors available back in 2005, and indicates that about 23\% of all current descriptors have been introduced during the last fifteen years.
% 2020 desciptors 29640
% 2005 desciptors 22997

\begin{table}
\caption{The distribution of the 6,915 new descriptors (2006 - 2020) into provenance codes. 
The total per category can be lower than the sum of distinct type counts as the types are not mutually exclusive.}
\label{tab:distribution}       % Give a unique label
\centering
\begin{tabular}{L{0.01\linewidth}L{0.20\linewidth}R{0.06\linewidth}R{0.06\linewidth}R{0.15\linewidth}|R{0.08\linewidth}}
\multicolumn{2}{l}{}       					& \multicolumn{3}{c}{\textbf{Prov. Category}}        			&		\\
\multicolumn{2}{l}{}  						& \textbf{1.}       	& \textbf{2.}      	& \textbf{3.}  			&	\textbf{Total}		\\
\multicolumn{2}{l}{\textbf{Prov. Type}}
& \textbf{Old con.} 	& \textbf{Old SCR} 	& \textbf{New PI con.} 	&	\textbf{/type}		\\
% \textbf{.0}   & \textbf{Emersion}    		&	 -           		& -       			& -       			& 1720 		\\
\textbf{.1}   & \textbf{Succession}  		&	 21	         		& 12	     		& 84	      			& 117 		\\
\textbf{.2}   & \textbf{Subdivision} 		&	 276         		& 967     			& 1,603       			& 2,846 		\\
\textbf{.3}   & \textbf{Submersion}  		&	 47	         		& 535     			& 506	       			& 1,088  		\\
\textbf{.4}   & \textbf{Overtopping} 		&	 24	         		& 7	     			& 91	       			& 122 		\\
\textbf{.5}   & \textbf{Detachment}  		&	 151         		& 364      			& 1,313            		& 1,828 		\\ 
\hline
\multicolumn{2}{c}{\textbf{Total/category}} 		&    519				& 1,616				& 3,060					& \textbf{5,195} \\ 
\multicolumn{6}{l}{The total for \textit{category 4}, Emersion (4.0), is 1,720.}  \\
% \multicolumn{6}{r}{*Total /category can be lower than the sum across all types, because types are not mutually exclusive.} 
\end{tabular}
\end{table}

\begin{figure*}
\center
\includegraphics[width=0.9\textwidth]{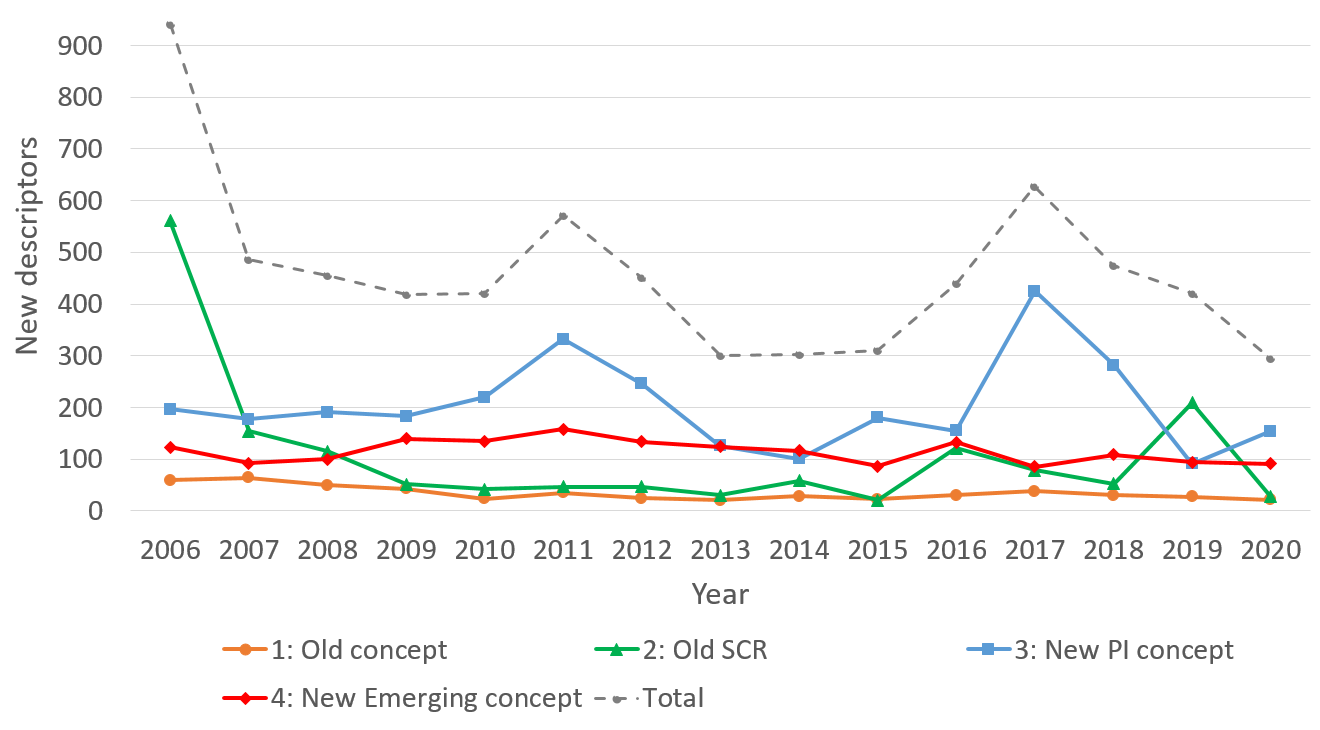}
\caption{Frequency of provenance categories for new descriptors, per year of introduction. }
\label{fig:categories_line}       % Give a unique label
\end{figure*}

The new descriptors introduced for new concepts that have been implicitly covered in their \textit{version 0} by old descriptors (\textit{category 3}) is the most frequent provenance category, accounting for about 44\% of all new descriptors. 
New descriptors for old concepts that have been explicitly covered in previous versions account for about 31\% of all new descriptors, with the majority of cases covered by SCRs (\textit{category 2}, $\sim$23\%) and only a small portion having a previous status as a subordinate concept (\textit{category 1}, $\sim$8\%).
This suggests that new descriptors for concepts already covered explicitly by older descriptors are primarily added for promoting SCRs to descriptors (\textit{category 2}), rather than for promoting subordinate concepts restructuring old descriptors (\textit{category 1}).

On the other hand, new descriptors for emerging concepts (\textit{category 4}), which are totally new for the MeSH vocabulary, account for 25\% of all new descriptors. 
This relatively low frequency of \textit{Emersion} suggests that in most cases new descriptors are linked to domain entities that are already covered by other descriptors either implicitly (\textit{category 3}) or explicitly (\textit{category 1} and \textit{2}). 
Therefore, the new conceptual entities that are very often introduced (\textit{category 3} and \textit{4} account for 69\% of new descriptors) are not completely novel, but they usually offer dedicated descriptors to known concepts (\textit{category 3}).
% These results suggest that a main force driving the extension of MeSH with new descriptors is the need to cover new conceptual entities explicitly, that were previously usually covered implicitly (\textit{category 3}) only or were not covered at all (\textit{category 4}).

% 	    3	    1   	2   	4   	Total
% AVG	204.0	34.6	107.7	114.7	461.0
% STDV	85.5	13.2	131.5	22.0	158.5

Furthermore, the annual distribution of new descriptor categories, shown in Fig.~\ref{fig:categories_line}, confirms the consistently high frequency of \textit{categories 3} and \textit{4} throughout the years.
In particular, both the introduction of descriptors for new PI concepts and new emerging concepts accounts for at least around 100 cases annually for the whole period of study. 
However, \textit{category 4} is more stable around its mean value (AVG) of almost 115 cases per year, with a standard deviation (SD) of 22 cases, whereas \textit{category 3} presents more variation around its mean of 204 cases (SD $\sim$85 cases), reaching up to 300 and 400 cases in certain years.

On the other hand, the promotion of existing SCRs into descriptors (\textit{category 2}) seems the less predictable category with an AVG around 108 and a SD of around 131 cases per year. In particular, in certain years (e.g. 2006, 2019) there seems to be a surge of such cases, while in others the number is much smaller. 
% This seems reasonable, as SCR promotion largely depends not only on the number of citations available for each SCR but also on the resources allocated for this task at NLM~\footnote{Cho, Dan-Sung (NIH/NLM) personal communication}.  
Finally, the evolution of existing subordinate MeSH concepts into independent descriptors (\textit{category 1}) seems the least frequent and the most stable category with an AVG of around 35 and a SD of around 13 new descriptors per year. 

\begin{figure*}
\center
\includegraphics[width=0.9\textwidth]{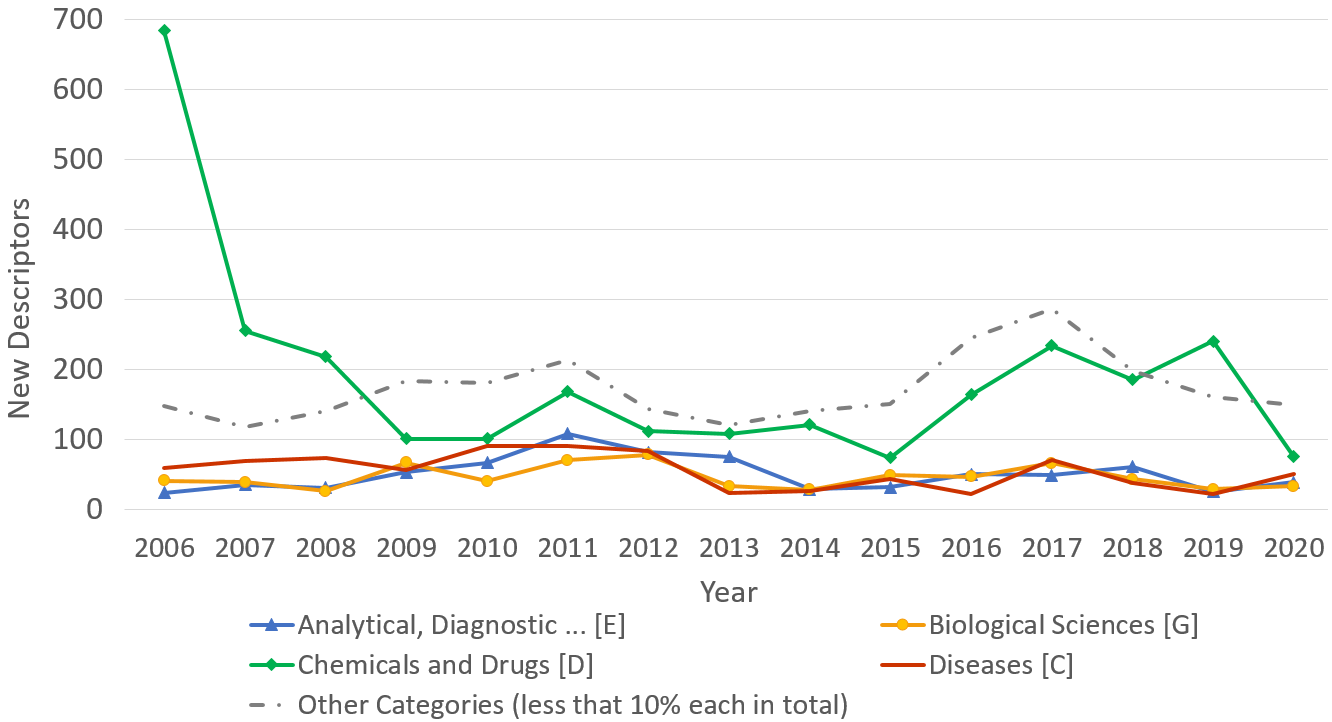}
\caption{Frequency of MeSH categories for new descriptors, per year of introduction. The four MeSH categories accounting for at least 10\% of new descriptors each are presented independently. The remaining twelve cases, that have an overall frequency of less than 10\% of new descriptors each, are collectively presented as ``Other Categories''.}
\label{fig:MeSH_categories_line}       % Give a unique label
\end{figure*}

The extreme peak of more than 900 new descriptors observed in 2006, may be the result of an effort at NLM to restructure descriptors for chemicals that combined meanings for activity and structure. 
This effort, which has been spanning across many years, was continued in 2006.\footnote{\url{https://www.nlm.nih.gov/pubs/techbull/nd05/nd05_2006_MeSH.html}}  
In addition, promoting SCRs to descriptors was particularly encouraged this year in NLM\footnote{Cho, Dan-Sung (NIH/NLM) personal communication}, which is in agreement with the fact that this peak seems to be almost exclusively attributed to promoted SCRs (\textit{category 2}), which are known to represent mainly chemicals.
This is also confirmed by the distribution of new descriptors into MeSH categories (Fig.~\ref{fig:MeSH_categories_line}), as 73\% of the new descriptors introduced in 2006 belong to ``Chemicals and Drugs'' (D). This relative frequency for 2006 far exceeds the overall relative frequency of category D for the whole period considered, that is around 41\%.   

Two less extreme peaks are also observed in 2011 and 2017, with the introduction of about 600 new descriptors each. In contrast to the 2006 peak, these ones seem to be primarily attributed in \textit{category 3} cases, as other categories present frequencies close to the ones of the adjacent years. 
In addition, the distribution of the corresponding new descriptors into MeSH categories suggests that, though the chemicals category D has relatively high frequencies these years, other MeSH categories also have a considerable contribution to these peaks.
% In particular, the percentage of category D seems to be lower in these peaks, than overall during the whole period. 
In other words, these peaks of new descriptors for new PI concepts (\textit{category 3}) seem to be more evenly distributed across MeSH categories, than the 2006 peak of \textit{category 2} cases.

For 2011, this is in agreement with a focus in MeSH on projects related to the categories ``Biological Sciences'' (G) and ``Analytical, Diagnostic and Therapeutic Techniques, and Equipment'' (E) in MeSH.\footnote{Cho, Dan-Sung (NIH/NLM) personal communication} 
The peak of 2017, on the other hand, seems to be affected by the ``MeSH Protein Project''\footnote{\url{https://www.nlm.nih.gov/pubs/techbull/nd16/nd16_mesh.html}}, as part of which, almost 290 new descriptors were added. 
The aim of this project was to achieve alignment of gene families, as described by the Human Genome Nomenclature Committee (HGNC), with protein classes in MeSH.
In addition, more new descriptors than usual are introduced in 2017 for some less frequent MeSH categories, such as ``Health Care'' (M) and ``Persons'' (N).   

Regarding the provenance types of new descriptors, \textit{Subdivision} (.2) is the most common case (41\%), followed by \textit{Detachment} (.5, 26\%) and \textit{Emersion} (.0, 25\%). \textit{Submersion} has also a considerable frequency of 16\%, but \textit{Succession} (.1) and \textit{Overtopping} (.4) are quite scarce, accounting for about 2\% each.
This distribution seems to be in agreement with the expected low frequency of new descriptors being broader of their PHs (\textit{Overtopping}) or having their PHs removed from the vocabulary (\textit{Succession}).
However, the frequency of new descriptors that are no longer covered by any of their PHs (\textit{Detachment}) seems quite notable, representing 35\% of non-emerging new descriptors (\textit{categories 1, 2} and \textit{3}). 
This implies that the addition of dedicated descriptors for concepts that used to be covered by older descriptors (PHs), often serves the removal of these subordinate, supplementary, or implicitly covered concepts from these PHs, improving the specificity of the latter. 
% this also supports the enrichment of unrelated descriptors with this topic. however, this enrichment is also supported by any type, when combined with n additional tree path.     

\begin{figure*}
\center
\includegraphics[width=0.85\textwidth]{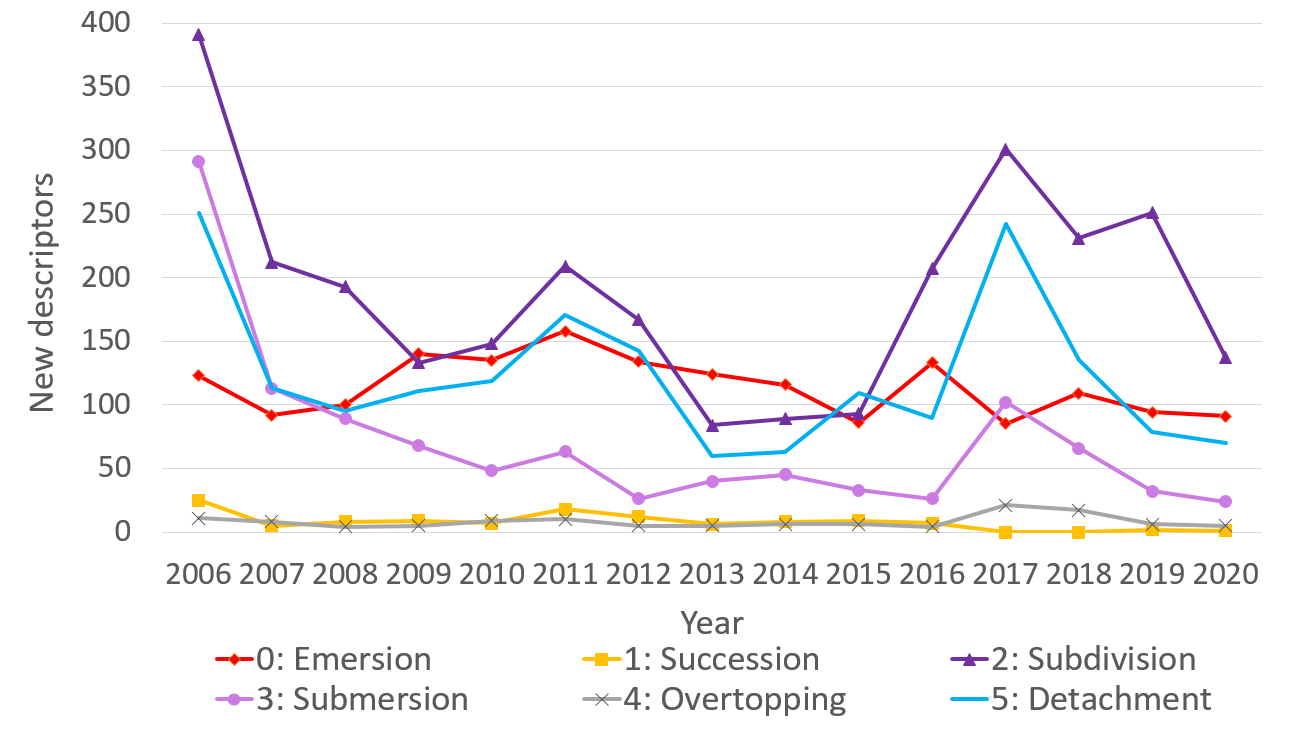}
\caption{Frequency of provenance types for new descriptors, per year of introduction. }
\label{fig:types_line}       % Give a unique label
\end{figure*}

On the other hand, the majority of new descriptors appear to be still covered by their PHs, offering subtopics to the latter. 
In particular, about 55\% of all the new descriptors have at least one \textit{ancestor} in their PHs, that is they belong to \textit{Subdivision} or \textit{Submersion} cases, with the last being far less frequent as expected (16\%).
This suggests that only half of the new descriptors end up as descendants of their PHs.
However, focusing on the 5,195 non-emerging new descriptors, that actually have at least one PH (\textit{categories 1, 2} and \textit{3}), this relative frequency increases to 73\%, with \textit{Subdivision} accounting for 55\% of the cases and \textit{Submersion} for only 21\% of them.
This is in agreement with the expected evolution of the topic vocabulary towards more fine-grained descriptors. The latter support more precise topic annotations and retrieval, especially when more documents are accumulated for some descriptors during the years. 

Figure~\ref{fig:types_line} presents the annual distribution of new descriptors into provenance types.
Despite annual fluctuations, there seems to be a clear separation of the frequent types (\textit{Emersion}, \textit{Subdivision}, and \textit{Detachment}), from the infrequent ones (\textit{Succession} and \textit{Overtopping}) throughout the period of study. 
Finally, the \textit{Submersion} type seems to fall in-between the two groups. 
In addition, it seems that the infrequent types of \textit{Succession} and \textit{Overtopping} vary the least through the years (SD 7 and 5 respectively).
The more frequent types of \textit{Subdivision}, \textit{Detachment} and \textit{Submersion} seem to be the less predictable (SD 81, 56 and 66 respectively), whereas the trivial type of \textit{Emersion}, though quite frequent as well, appears to be relatively stable, as already noticed for \textit{category 4}.

As with MeSH categories, the surge of cases in certain years is not evenly distributed across all provenance types. 
Although the representation of all provenance types appears to be close to their overall relative frequency in the peak of 2011, this is not always the case.
In 2006, \textit{Submersion} seems to be over-represented, accounting for 31\% of the cases, which is more than double its overall relative frequency for the period of study (16\%). This could be related to the complex organization of chemical SCRs into groups and subgroups. 
For example, ``Receptors, Scavenger'' as well as the six classes of them (``Scavenger Receptors, Class A'' etc) used to be SCRs indexed under ``Receptors, Immunologic'' until their promotion into descriptors in 2006. Although ``Receptors, Scavenger'' was added as a child (2.2) to their PH ``Receptors, Immunologic'', the six classes were added as children of ``Receptors, Scavenger'', hence more distant descendants of ``Receptors, Immunologic'' (2.3). 

On the other hand, \textit{Detachment} seems to be over-represented in the peak of 2017, accounting for 39\% of the new descriptors, whereas its overall relative frequency for the whole period is 26\%.
Some of these \textit{Detachment} cases are new descriptors for protein domains or motifs detached from the corresponding protein descriptors, which can be related to the ``MeSH Protein Project''. For example, the new descriptor ``Methyl CpG Binding Domain'' detached from its PH ``DNA-Binding Proteins''.
In addition, several new descriptors in the MeSH categories ``Health Care'' (M) and ``Persons'' (N) appear to represent medical professions detached from the corresponding medical domains. For example, the new descriptor ``Nephrologists'' was detached from its PH ``Nephrology''.

\begin{figure*}
\center
\includegraphics[width=0.9\textwidth]{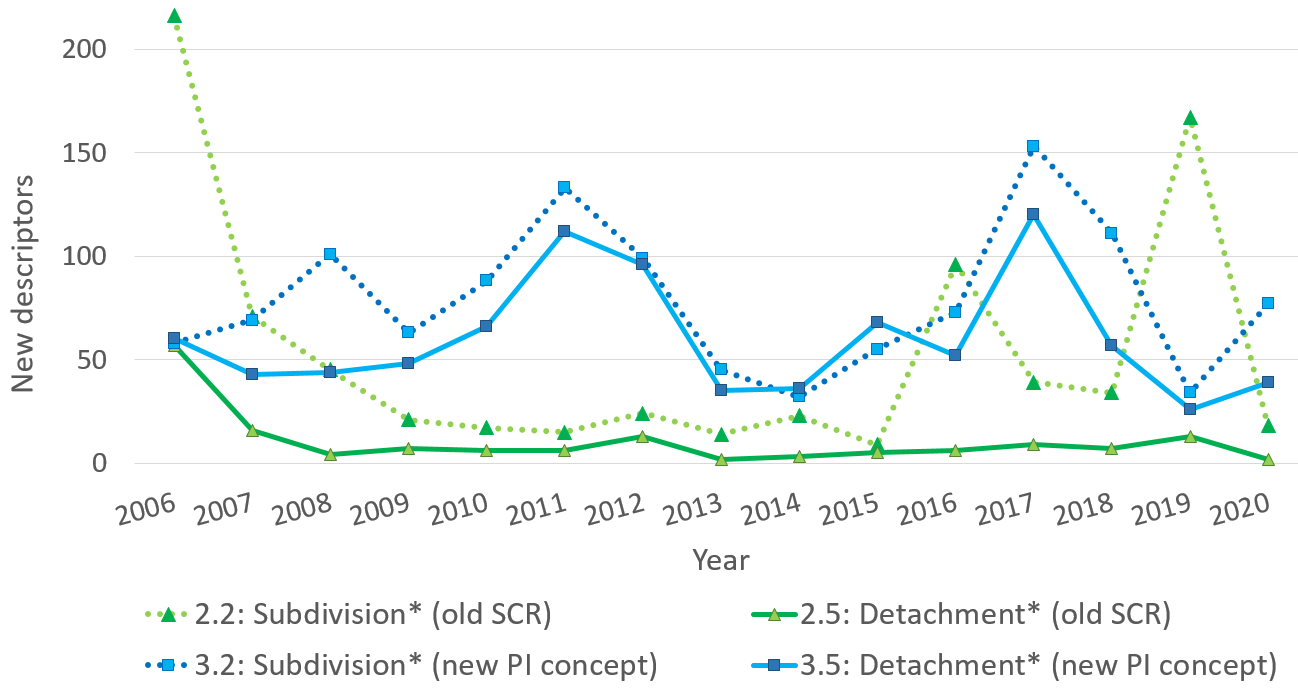}
\caption{Frequency of type \textit{Subdivision} (.2) and \textit{Detachment} (.5) in new descriptors introduced during the last fifteen years, per provenance category. The asterisk (*) denotes that only descriptors with a single type are considered, excluding descriptors combining more than one type.}
\label{fig:types_2_6_line}       % Give a unique label
\end{figure*}

Some of the types, in particular \textit{Subdivision} (.2) and \textit{Detachment} (.5), seem to be correlated in the way they increase or decrease over the years. 
It would, therefore, be of interest to investigate whether the correlation of their annual frequencies observed in Fig.~\ref{fig:types_line} should be attributed to the addition of descriptors that exhibit both these provenance types simultaneously.
This is mainly possible in \textit{category 2} and \textit{category 3} where the availability of multiple PHs for a new descriptor can lead to multiple provenance codes. 
In practice, however, new descriptors with multiple provenance codes are not very common, representing almost 17\% of all new descriptors in these two categories.
% 770 with >1 p.types / 4676 cat 2+3
% 1388 with >1 PHs / 4676 cat 2+3

Focusing on the majority of new descriptors that have a single provenance type, we compare the annual frequencies of the \textit{Subdivision} (.2) and \textit{Detachment} (.5) (Fig.~\ref{fig:types_2_6_line}). 
The correlation of the frequencies seems to be preserved in the frequent \textit{category 3} (blue lines with square markers). In other words, even when looking at distinct new descriptors that share no common provenance types, \textit{Subdivision} (3.2) and \textit{Detachment} (3.5) seem to fluctuate in the same way across the years. 
For \textit{category 2} on the other hand (green lines with triangle markers), \textit{Detachment} (2.5) doesn't seem to keep-up with \textit{Subdivision} (2.2) which presents some high peaks (2006, 2016, 2019).
This is reasonable, as the link of the new descriptors to their PHs is stronger in \textit{category 2}, which is based on explicit coverage, compared to \textit{category 3} where the PHs used to cover the new descriptors only implicitly. 

It appears that in \textit{category 3}, the amounts of new descriptors that are added as children of their PHs are usually comparable to the ones that are detached from their PHs.
This observation could be the effect of an internal procedure in the maintenance of MeSH and may warrant further investigation. 
On the other hand, the frequency of emerging descriptors without any PHs (Emersion 4.0) (Fig.~\ref{fig:types_line}) exhibits fluctuations that are not particularly correlated to the other frequent types of provenance. This suggests that the addition of descriptors with totally new preferred concepts forms a distinct subset of the new descriptors added each year.

\section{Conclusion and Future Work}
\label{sec:Conclusion}

In this work, we proposed a novel conceptual framework for organizing and studying the conceptual provenance of new descriptors in the Medical Subject Headings (MeSH) Hierarchy.
In particular, we defined the notion of the previous host (PH), as a descriptor covering the main topic of a new descriptor prior to its introduction, and suggested an approach to identify such PHs for a new descriptor. 
Then, based on the current relationship of the descriptor with its PHs we also defined a set of provenance types and codes.
In addition, we developed an open-source computational process for the automated extraction, annotation, and analysis of new descriptors, using the raw files of different versions of MeSH as distributed by NLM. 
Employing this approach, we investigated the conceptual provenance of new MeSH descriptors for the period 2006-2020.

The results reveal that about 115 new descriptors for emerging concepts (\textit{category 4}) are introduced each year quite steadily.
These descriptors represent about 25\% of all new descriptors of the study period, indicating that the majority of the new descriptors cover non-emerging domain concepts that are not really new for the MeSH thesaurus.
Less than half of these non-emerging concepts were explicitly covered in MeSH prior to the introduction of dedicated descriptors for them (\textit{category 1} and \textit{category 2}). 
% 44 / 100-25 = 59 cat 3 in non-emerging
The majority of non-emerging concepts, though not explicitly included in older versions of MeSH, used to be indexed under specific older descriptors (PHs) that covered their meaning implicitly (\textit{category 3}).

This suggests that the main force which is consistently driving the extension of MeSH during this period is the need to explicitly cover more conceptual entities.
Namely, a stable annual amount of new emerging concepts (\textit{category 4}) and a similar or greater amount of new PI concepts (\textit{category 3}), that used to be implicitly covered by MeSH.
The need to introduce descriptors for reorganizing concepts that are already explicitly covered (\textit{category 1} and \textit{category 2}) appears to be auxiliary, with low amounts of new descriptors for most years.
However, in certain years, we also observed a surge in the promotion of existing SCRs into descriptors (\textit{category 3}), particularly for chemicals.
Such surges in \textit{category 2} and \textit{category 3}, seem to be related with internal MeSH projects and resource allocation in NLM.

In addition, the results on conceptual provenance types reveal that more than 70\% of all non-emerging new descriptors (\textit{categories 1, 2} and \textit{3}) become subtopics of their PHs' topics. 
That is, they remain under the coverage of the latter, usually as children of them (.2, \textit{Subdivision}) and less often as more distant descendants (.3, \textit{Submersion}). 
However, the amount of new descriptors that are detached from their PHs (.5, \textit{Detachment}) is also considerable, particularly for implicit PHs (\textit{category 3}).
These observations suggest that the extension of MeSH primarily serves the need to enrich the MeSH thesaurus with more detailed subtopics, supporting the annotation of articles with new fine-grained topic labels.   
Nevertheless, it appears that a notable amount of new descriptors also serve to rid the PHs of some implicitly covered topics, rendering the PHs more precise as well.  

This grouping can be particularly useful for improving semantic indexing models for new descriptors. 
For example, the articles annotated with their PHs can be a source of weakly labeled data for topical annotations.  
In addition, the provenance types can provide indications for the prevalence of such weak labels. In the case of \textit{Detachment} for example, we may expect that only a small part of the articles annotated with the PHs will be relevant to the new descriptor.
% On the other hand, provenance categories and types can also be useful for the retrospective update of old annotations, to achieve more homogeneity among articles annotated in different periods. 
% For example, once we have a sufficiently good model for a new detached descriptor, we could 
% , in 
In the case of new descriptors for new emerging concepts (\textit{category 4}), on the other hand, Zero-Shot Learning approaches may be more appropriate as no PHs are available as a source of weak labels.

% \todo[inline]{up to here}

Although our findings primarily provide insight to researchers working with MeSH, we believe that the proposed approach is of more general interest. In particular, it can be adapted for analyzing the extension dynamics of other similar topic hierarchies. In the lack of appropriate fields, such as subordinate concepts, PMN and PI, for identifying explicit or implicit PHs in other hierarchies, one could explore term-matching approaches in different hierarchy versions.  
The annotations of conceptual provenance produced by the proposed method capture the hierarchical relationship of a new topic with the topics that were previously used in its place. Such information can be used to characterise and group the topics, facilitating the process of maintaining topic hierarchies.

Our future plans include the investigation of further uses of the provenance information provided by the proposed method. In particular, we are examining whether new descriptors with the same provenance category, types or codes, present similarities that can be exploited in the semantic indexing of documents with newly introduced labels. 
Additionally, we are looking into the use of provenance information for predicting ontological expansion. 
Last but not least, we would like to explore the use of the conceptual framework and computational procedures for tasks related to the maintenance of the hierarchy itself, such as identifying special cases and inconsistencies in textual descriptive fields.

\begin{acknowledgements}
This research work was supported by the Hellenic Foundation for Research and Innovation (HFRI) under the HFRI Ph.D. Fellowship grant (Fellowship Number: 697).
We are grateful to James Mork and Dan-Sung Cho from the National Library of Medicine (NLM) for kindly providing valuable feedback on this work.
\end{acknowledgements}


\bibliographystyle{spmpsci}      % mathematics and physical sciences

\bibliography{main}

% % Non-BibTeX users please use
% \begin{thebibliography}{}
% %
% % and use \bibitem to create references. Consult the Instructions
% % for authors for reference list style.
% % Format for Journal Reference
% Author, Article title, Journal, Volume, page numbers (year)
% Author, Book title, page numbers. Publisher, place (year)
% \end{thebibliography}

% \section{Appendix}
% % \todo[]{Add as electronic supplementary material}

\end{document}